\documentclass[amsmath,aps,twocolumn]{revtex4-1}
\usepackage{graphicx}
\usepackage{times}
\usepackage{slashed}
\usepackage{bm}
\usepackage{amsmath}
\usepackage{amssymb}
\usepackage{multirow}

\begin{document} 

\title{Theoretical prerequisites for the upcoming generation of precision spectroscopic experiments}

\author{Dmitry Solovyev$^{1,2}$} 
\email[E-mail:]{d.solovyev@spbu.ru.}
\author{Timur Zalialiutdinov$^{1,2}$, Aleksei Anikin$^{1,3}$ and Leonti Labzowsky$^{1,2}$}

\affiliation{$^{1}$Department of Physics, St. Petersburg State University, St. Petersburg, 198504, Russia
\\
$^{2}$Petersburg Nuclear Physics Institute named by B.P. Konstantinov of National Research Centre "Kurchatov Institut", St. Petersburg, Gatchina, 188300, Russia
\\
$^{3}$D.I. Mendeleev Institute for Metrology, St. Petersburg, 190005, Russia}

\date{\today}

\begin{abstract}

Modern resonant spectroscopic experiments to measure transition frequencies in atoms have reached a level where a meticulous description of all aspects of the processes under study has become obligatory. The precision achieved in the experiments of A. Beyer, et al., Science 358, 79 (2017), has led to the fact that the determination of the transition frequency based on measured data is significantly refined by theoretical treatment of the observed spectral line profile. As it was predicted theoretically, a great impact of effects arising beyond the resonance approximation was found experimentally. These findings marked the beginning of the upcoming epoch in the resonant atomic spectroscopy when many commonly understood ideas became invalid. For example, the atomic transition may be characterized by several different but equally acceptable frequencies. Furthermore, we show that the picture becomes even more complicated when the observed spectral line profile is ”identified” with one of the processes - emission or absorption. Precise determination of the transition frequency requires a description of the absorption line profile inseparable from the emission process and vice versa. The theoretical aspects discussed in this work provide prerequisites for more accurate and yet simpler experiments than those reported in Science 358, 79 (2017). Implementing the new physics expected in atomic resonance spectroscopy in the near future beyond the resonance approximation is unfeasible without resolving these issues.
\end{abstract}

\maketitle 

\section{Introduction}

At present, experimental progress ascertained that the spectral line profile can be studied in detail as a precisely measurable quantity. The theory of natural line profile in atomic physics was developed since the pioner works \cite{Weisskopf1930} within Quantum Mechanics (QM) and \cite{Low} within Quantum Electrodynamics (QED). An extended QED theory of the line profile for many-electron atoms and highly charged ions (HCI) was described in \cite{ANDREEV2008135}, see also \cite{ZALIALIUTDINOV20181}. The development of the line profile theory is closely related to experimental advances in measuring the transition frequency in the hydrogen atom \cite{PhysRevLett.84.5496,PhysRevLett.110.230801}, where the $1s-2s$ two-photon transition was measured with an accuracy of about $10^{-15}$ relative magnitude. Such experiments have stimulated interest in theoretical studies of effects beyond the resonance approximation. The most important consequences in the line profile theory were found in the form of corrections to the transition frequency arising due to nonresonant terms in the photon scattering cross section \cite{2001,Jent-NR}. These corrections (called nonresonant, NR) expressly demonstrate the breach of the resonance approximation. The largest contribution originates from the states most close by energy to the resonant one. 

Even larger contributions come from the interference between the resonant and nonresonant states. However, in the total cross-sections, the interference terms survive only for the states with the same symmetry. Such states are rather rarely located close by energy in atomic spectra. Therefore, an important step that initiated further research was made in \cite{Jent-NR}, where NR corrections were considered in the differential cross-sections, most offen observed in experiments, and the distorted profile used later in \cite{Science} was derived. In this case interference NR corrections for the states with different symmetries contribute to the line profile and namely these corrections, called Quantum Interference Effects (QIE), are the main topic of  theoretical and experimental studies \cite{PhysRevLett.98.203003,PhysRevA.91.062506,PhysRevA.97.022510,PhysRevA.92.022514,https://doi.org/10.1002/andp.201900044,Matveev_2019}. 

In the general case, the QI effect depends on the angles between the vectors characterizing the emitted and absorbed photons (photon polarization, propagation direction or their combination). This, in turn, opens the possibility to avoid the NR-shift of the transition frequency by the choice of such angle when the QIE is equal to zero ("magic angle"). Any NR corrections make the line profile asymmetric (by mere construction), while NR corrections (though much smaller) to the total cross section do not depend on "experimental geometry" and, therefore, the corresponding asymmetry is always present. Basing on the ever available asymmetry of the line profile it was concluded in \cite{2001,Jent-NR,PhysRevA.65.054502} that NR corrections set a limit to the accuracy of the transition frequency determination.

It was recently argued in \cite{Science} that even for an observed asymmetric line profile, the transition frequency can be unambiguously determined giving an "invariant" that is independent of the experimental conditions. This can be achieved by subtracting the asymmetric part of the line profile. The subtraction procedure described in \cite{Science} was grounded in the Fano-Voit profile, since the QIE was first discussed by Fano \cite{Fano} and the Voit line profile parameterization was used. The concept of symmetrization of the observed line profile \cite{Science} is intended to determine the frequency of a particular transition with a number of significant digits far in excess of the NR corrections. However, the most specific for nonresonant corrections is their dependence on the process used in the experiment, and, therefore, NR corrections are an inherent part of the experimental conditions. Since the nonresonant corrections are the result of going beyond the resonant approximation, the transition frequency can be uniquely determined only within the resonant approximation. 

For the more detailed investigation of the problem we have to recall a fundamental requirement of the theory: the scattering process has to start and end with a stable state \cite{Bjorken}. In the traditional atomic resonant spectroscopy this requirement was ignored and the transitions between two unstable states (cascade transitions) were commonly observed, see, for example, \cite{deBov0,deB-1,PhysRevA.90.012512,PhysRevA.95.052503,Fleurbaey}. The main goal of this communication is to show that with the growth of the accuracy for spectroscopic measurements this requirement becomes obligatory. An excellent example which helps to justify this statement is the recently published paper \cite{Science} on the superaccurate measurement of $2s-4p$ transition frequency in hydrogen atom. The final state $4p$ in this transition is unstable. Its decay occurs among other channels also via cascades. In spite of the fact that the cascades represent fractions from the dominant decay to the ground state, their analysis for nonresonant effects and, as a consequence, the asymmetry of the line profile is a matter of principal importance. It should be emphasized that the experiment \cite{Science} can be considered the first where NR effects were observed. The emergence of quantum interference effects in the cascade (QIEc) emission process and their engagement in the absorption line profile is discussed in the last part of the present paper.

Measuring the absorption transition frequency to an unstable state leads to a set of possible situations beyond the resonance approximation, each of which is consistent with certain experimental conditions. For example, one possible way to determine the transition frequency corresponds to the case when all emitted photons are registered \cite{Science}, where due to the processing of the experimental data, the frequency was defined as the "line center".
Possible experiments where the excited state decays into fixed allowed lower states were discussed in \cite{Solovyev2020}. In this case the decay to a particular state leads to an asymmetry different from that revealed in \cite{Science}. Then the determination of the "line center" should be made by choosing a different asymmetry parameter. The transition frequency can be determined for any particular photon scattering channel with the same right as one defined in \cite{Science}.

This paper has several goals. One, as stated above, refers to the need to account for cascade radiation, to determine the absorption frequency. Another one is to demonstrate the existence of some "invariant" frequency (with respect to the experimental conditions and within the experimental error bars) for every possible definition of transition frequency beyond the resonance approximation. 

\section{Reconciliation among the various determinations of $2s-4p$ transition frequency}
\label{sec-fixed}

In this section, we discuss issues related to determining the transition frequency in spectroscopic experiments where the line profile is measured with high precision. These aspects are directly associated with the definition of optical standards and the subsequent calculation of the fundamental physical constants. Hereinafter, by the frequency standard ("frequency invariant") we will understand the value that can be reproduced in experiments repeating exactly the same furnishing. There are several obvious ways to extract the transition frequency from the observed line profile. The one of them, commonly accepted in theoretical analysis, is represented by the extremum condition and the definition of the transition frequency as the most probable, see, e.g., \cite{Jent-NR,PhysRevA.92.022514,PhysRevA.97.022510,PhysRevA.92.062506,PhysRevA.103.022833,Zalialiutdinov2021}. Another way corresponds to the determination of the "line center", see \cite{Science}. Obviously, these two recipes give the same result for a symmetric line profile. Consequently, within the resonance approximation both can be used to determine the frequency standard. Owing to asymmetry, which arises when nonresonant contributions to the photon scattering amplitude are taken into account, the maximum and the "line center" in the general case may not coincide. So, going beyond the resonance approximation, defining a frequency standard meets obvious obstacles. i) Using the extremum condition to determine the transition frequency, the asymmetry of the line shape can be accounted for as an additional frequency shift arising from the nonresonant terms in the scattering amplitude. This, however, can be effectively recognized with a theoretical description of the process used in the experiment. ii) For an asymmetric line profile, the "line center" cannot be uniquely determined, so a symmetrization procedure is required \cite{Science}. This was the reason for the conclusion that there is no unambiguous definition of the transition frequency beyond the resonance approximation \cite{Labzowsky1994,2001,Jent-NR}.

Both mentioned concepts are correlated by the line profile model used to match the measured line profile, but in general can lead to different results in determining the transition frequency. Theoretical background shows that the line profile has a parameter $\omega_0$, which can be used as a frequency "invariant" (this parameter serves as the most probable value for the symmetric profile and correspondsto the eigenvalues of the Hamilton operator). Similarly, the symmetrization procedure for the observed profile should lead to an unambiguous definition of the "line center". By obtaining a symmetrical shape, the most probable and the "line center" should coincide. However, the coincidence of the definitions within the two concepts depends on how accurately the process has been estimated theoretically or the symmetrization procedure has been applied. For example, as was shown in \cite{Solovyev2020}, a photon scattering process with a fixed final state can be used for this purpose. Obviously, the symmetrization procedure should reproduce the transition frequency found in \cite{Science}. At the same time, the "line maximum" and "line center" definitions of the transition frequency may differ essentially. So, at first sight, a conclusion can be made that there is no unambiguous way to determine the "absolute frequency" for a given transition, i.e., a universal value, an invariant, which can be achieved for any experimental conditions. Below we discuss one evident way to bring the definitions of this invariant down to a single value, at least within the experimental accuracy, i.e., to the coinciding values of "line center" and "line maximum". 


Focusing on the definition of the transition frequency $2s-4p$, as in \cite{Science}, in this section we will limit ourselves to describing the QIE for the resonance state $4p$ only \cite{Jent-NR,Solovyev2020}, while the eligibility to separate the absorption profile from the total process (i.e. from emission) is discussed in the next part of the paper. The interfering pathways are given by the transitions $2s^{F=0}_{1/2}\rightarrow 4p^{F=1}_ {1/2}$ and $2s^{F=0}_{1/2}\rightarrow 4p^{F=1}_{3/2}$, and the final result, employing the "line maximum" definition, can be expressed as (see Appendix~\ref{SuppM2}):
\begin{eqnarray}
\label{1}
\omega_{\rm max} = \omega_0+\delta\omega_r
\nonumber
\\
\delta\omega_r = \frac{f_{\rm nr}}{f_{\rm res}}\frac{\Gamma^2_r}{4\Delta_r}.
\end{eqnarray}
Here $\omega_0$ can be defined as the difference of the eigenvalues of the total Hamiltonian with inclusion of relativistic, QED, etc. effects) and corresponds to the line center (maximum) of the symmetric line profile. We characterize the atomic state by the principal quantum number $n$, the orbital momentum $l$, the total angular momentum $j$, accounting for the electron spin, $s$, and the total atomic momentum $F$ due to the nuclear spin momentum, $I$. $\Gamma_r$ represents the natural level width of the resonant excited state $n_rl_r$, and $\Delta_r$ represents the corresponding fine structure interval. Considering the particular case of $2s^{F=0}_{1/2}\rightarrow 4p^{F=1}_{1/2}$ and $2s^{F=0}_{1/2}\rightarrow 4p^{F=1}_{3/2}$ transitions the numerical values for the level width $\Gamma_r=1.2941\times 10^7$ Hz and the fine structure interval $\Delta_r = E_{4p^{F=1}_{3/2}} - E_{4p^{F=1}_{1/2}} = 1367433.3$ kHz, see \cite{HorbHess}, can be used with a sufficient accuracy, giving $\delta\omega_r$ value up to four digits after the decimal point. The NR correction in Eq.~(\ref{1}) is the QIE given by the angular dependence, for example, between the polarization vector of the incident photon $\vec{e}_i$ and the direction vector of the emitted photon $\vec{\nu}_f$, with the amplitudes $f_{\rm nr}$, $f_{\rm res}$.

The concept of determining $\omega_{\rm max}$ from Eq.~(\ref{1}) implies the existence of a frequency invariant $\omega_0$, representing the most probable transition frequency value for a symmetric line profile.

The amplitudes $f_{\rm nr}$ and $f_{\rm res}$ are uniquely determined by the quantum numbers of the states involved in the process under study. If in the process of the frequency measurement only the emission of the outgoing photon is detected without fixing its frequency \cite{Science}, the summation over all the final states should be done as follows
\begin{eqnarray}
\label{2}
\delta\omega_r = \frac{\sum\limits_{n_f l_f j_f F_f} f_{\rm nr}}{\sum\limits_{n_f l_f j_f F_f}f_{\rm res}}\,\frac{\Gamma^2_r}{4\Delta_r}.
\end{eqnarray}
Otherwise, specific scattering channels are defined by the dependence on the set of quantum numbers $n_f l_f j_f F_f$ in $f_{\rm nr}$ and $f_{\rm res}$. A straightforward comparison of the expressions Eqs.~(\ref{1}), (\ref{2}) demonstrates the difference in approaches to determining the transition frequency. In particular, it is literally seen that the angular dependence in these expressions does not have to be the same.

For the process of detecting all outgoing photons, in \cite{Science} it was found that the observed line profiles for transitions $2s^{F=0}_{1/2}\rightarrow 4p^{F=1} _{1/2}$ ($\nu_{1/2}$) and $2s^{F=0}_{1/2}\rightarrow 4p^{F=1}_{3/2}$ ($\nu_{3/2}$) are asymmetric. The resulting asymmetry is exactly consistent with Eq.~(\ref{2}). The angular factor involved in the ratio $\sum\limits_{n_f l_f j_f F_f} f_{\rm nr}/\sum\limits_{n_f l_f j_f F_f}f_{\rm res}$ can be expressed through a second-order Legendre polynomial: $P_2(\cos\theta) = (1/4)(1 + 3\cos 2\theta)$. Solving the equation $P_2(\cos\theta)=0$ ($\theta=\widehat{(\vec{e}_i,\vec{\nu}_f)}$), one can find the magic angle $\theta = \pm \arccos(1/\sqrt{3})+\pi n$ (with an arbitrary integer $n$). Thus, the asymmetry of the observed line profiles can be avoided by appropriate choice of the angle. However, in the experiment \cite{Science}, a special procedure for subtracting the asymmetric part of the line profile (by introducing the asymmetry parameter into the modified Fano-Voigt contour) was used to determine the transition frequencies. 
 The determined values for $\nu_{1/2}$ and $\nu_{3/2}$ were then used to calculate the weighted average of the hyperfine centroid, $\nu_{2s-4p}$, corrected for the hyperfine shift (noted here as $\Delta_{\rm HFS}$):
\begin{eqnarray}
\label{3}
\nu_{2s-4p} = \frac{1}{3}\nu_{1/2} + \frac{2}{3}\nu_{3/2} - \Delta_{\rm HFS},
\\
\nonumber
\Delta_{\rm HFS} = 132552.092(75)\, {\rm kHz}.
\end{eqnarray}
As a result of this processing, the weighted average value $\nu_{2s-4p}$ ("line center") was standardized as the transition frequency, which was then used to determine the proton charge radius and the Rydberg constant. Although significant progress has been made in solving the "proton radius puzzle," the experiments \cite{Fleurbaey,ediss27002} do not eliminate the problem, see \cite{Pohl}. So, the question of accuracy and universality of determining the transition frequency remains relevant.

Turning to another possible experimental conditions of determining the transition frequency, the case when the final state is fixed can be considered. Accordingly, see Appendix~\ref{SuppM2}, this NR correction is defined by the expression Eq.~(\ref{1}) and does not depend on any angles between scattered photons \cite{Solovyev2020}. The obtained values for the partial channels are given in Table~\ref{tab:1}, where the $\omega_0$ values and the resonant state width were used as parameters determining the line profile. The corresponding nonresonant frequency shifts are also given. 
\begin{widetext}
\begin{center}
\begin{table}
\renewcommand{\arraystretch}{1.2}
\caption{Numerical values of the transition frequencies $\omega_{\rm max}$. The $\omega_0$ values, see \cite{HorbHess}, used in the calculations are shown in the second column. The third column shows the NR correction values Eq.~(\ref{1}) for specific scattering channels. The last column contains the $\omega_{\rm max}$ values, summed over the total momentum $F_f$ with weight $(2F_f+1)/(2j_f+1)(2I+1)$ and additional averaging factor $1/(2j_r+1)$, $\omega_{\rm av}$. The last column shows also $\nu_{1/2}$, $\nu_{3/2}$, borrowed from \cite{Science}. The very last row shows values obtained for $\nu_{2s-4p}$. 
 All values are given in kHz.}
\label{tab:1}
\begin{tabular*}{\textwidth}{@{\extracolsep{\fill}}l l l l l}
\hline
Transition & $\omega_0$ in kHz, see \cite{HorbHess} & $\delta\omega_r$ in kHz & $\omega_{\rm max}$ in kHz & $\omega_{\rm av}$ in kHz\\

\hline
\hline
 $2s^{F=0}_{1/2}\rightarrow 4p^{F=1}_{1/2}\rightarrow 1s_{1/2}^{F_f=0}$ & \multirow{2}{*}{$616520152558.5$} & $60.7127$ & $616520152619.2$ & \multirow{2}{*}{$616520152554.7$} \\
 $2s^{F=0}_{1/2}\rightarrow 4p^{F=1}_{1/2}\rightarrow 1s_{1/2}^{F_f=1}$ &  & $-30.3563$ & $616520152528.1$ &  \\
 
\hline
$2s^{F=0}_{1/2}\rightarrow 4p^{F=1}_{1/2}\rightarrow 2s_{1/2}^{F_f=0}$ & \multirow{2}{*}{$616520152558.5$} & $60.7127$ & $616520152619.2$ & \multirow{2}{*}{$616520152554.7$} \\
 $2s^{F=0}_{1/2}\rightarrow 4p^{F=1}_{1/2}\rightarrow 2s_{1/2}^{F_f=1}$ &  & $-30.3563$ & $616520152528.1$ &  \\

\hline
$2s^{F=0}_{1/2}\rightarrow 4p^{F=1}_{1/2}\rightarrow 3s_{1/2}^{F_f=0}$ & \multirow{2}{*}{$616520152558.5$} & $60.7127$ & $616520152619.2$ & \multirow{2}{*}{$616520152554.7$} \\
 $2s^{F=0}_{1/2}\rightarrow 4p^{F=1}_{1/2}\rightarrow 3s_{1/2}^{F_f=1}$ &  & $-30.3563$ & $616520152528.1$ &  \\
  
\hline
$2s^{F=0}_{1/2}\rightarrow 4p^{F=1}_{1/2}\rightarrow 3d_{3/2}^{F_f=1}$ & \multirow{2}{*}{$616520152558.5$} & $-30.3563$ & $616520152528.1$ & \multirow{2}{*}{$616520152554.7$} \\
 $2s^{F=0}_{1/2}\rightarrow 4p^{F=1}_{1/2}\rightarrow 3d_{3/2}^{F_f=2}$ &  & $6.0713$ & $616520152564.6$ &  \\
 
 \hline
$\nu_{1/2}$, rms & $616520152558.5$ & & $616520152566.8$ & $616520152554.7$\\
$\nu_{1/2}$, Ref.~\cite{Science} & & & & $616520152555.1(3.0)$\\

\hline
\hline

 $2s^{F=0}_{1/2}\rightarrow 4p^{F=1}_{3/2}\rightarrow 1s_{1/2}^{F_f=0}$ & \multirow{2}{*}{$616521519991.8$} & $-15.1782$ & $616521519976.6$ & \multirow{2}{*}{$616521519996.5$} \\
 $2s^{F=0}_{1/2}\rightarrow 4p^{F=1}_{3/2}\rightarrow 1s_{1/2}^{F_f=1}$ &  & $30.3563$ & $616521520022.2$ &  \\
 
\hline
$2s^{F=0}_{1/2}\rightarrow 4p^{F=1}_{3/2}\rightarrow 2s_{1/2}^{F_f=0}$ & \multirow{2}{*}{$616521519991.8$} & $-15.1782$ & $616521519976.6$ & \multirow{2}{*}{$616521519996.5$} \\
 $2s^{F=0}_{1/2}\rightarrow 4p^{F=1}_{3/2}\rightarrow 2s_{1/2}^{F_f=1}$ &  & $30.3563$ & $616521520022.2$ &  \\

\hline
$2s^{F=0}_{1/2}\rightarrow 4p^{F=1}_{3/2}\rightarrow 3s_{1/2}^{F_f=0}$ & \multirow{2}{*}{$616521519991.8$} & $-15.1782$ & $616521519976.6$ & \multirow{2}{*}{$616521519996.5$} \\
 $2s^{F=0}_{1/2}\rightarrow 4p^{F=1}_{3/2}\rightarrow 3s_{1/2}^{F_f=1}$ &  & $30.3563$ & $616521520022.2$ &  \\
  
\hline
$2s^{F=0}_{1/2}\rightarrow 4p^{F=1}_{3/2}\rightarrow 3d_{3/2}^{F_f=1}$ & \multirow{2}{*}{$616521519991.8$} & $30.3563$ & $616521520022.2$ & \multirow{2}{*}{$616521519970.9$} \\
 $2s^{F=0}_{1/2}\rightarrow 4p^{F=1}_{3/2}\rightarrow 3d_{3/2}^{F_f=2}$ &  & $-151.7819$ & $616521519840.0$ &  \\
 
 \hline
$\nu_{3/2}$, rms & $616521519991.8$ & & $616521519982.3$ & $616521519990.1$\\
$\nu_{3/2}$, Ref.~\cite{Science} & & & & $616521519990.8(3.0)$\\
\hline
\hline
$\nu_{2s-4p}$, rms & $616520931628.6$ & & $616520931625.1$ & $616520931626.2$\\
$\nu_{2s-4p}$, Ref.~\cite{Science} & & & & $616520931626.8(3.0)$\\
\hline
\hline

\end{tabular*}
\end{table}
\end{center}
\end{widetext}

In \cite{Solovyev2020} it is stated that the obtained values can be used as a transition frequency with the same rights as the value determined in \cite{Science}. Moreover, each of them can be considered as a frequency standard if the experiment provides accurate registration of the state into which decay occurs. By subtracting $\delta\omega_r$ they can be reduced to $\omega_0$ \cite{HorbHess}, i.e. to the "invariant" frequency. Thus, the definitions of the transition frequency as "line center" and "line maximum" result in a rather significant difference of about 3 kHz, although lying far beyond the resonant approximation and within the experimental error. Below we show how this seeming contradiction can be resolved.

To compare the values $\omega_{\rm max}$ with those published in \cite{Science}, an averaging procedure should be applied to find the weighted average hyperfine centroid. For example, using the factor $(2F_f+1)/(2j_f+1)(2I+1)$ for final hyperfine sublevels and the averaging factor $1/(2j_r+1)$ for the resonant state (pseudo initial state for the emission process), the averaged centroid is
\begin{eqnarray}
\label{av0}
\omega_{\rm av} =  \omega_0+
\frac{1}{2j_r+1}\sum\limits_{F_f,j_f}
\frac{2F_f+1}{(2j_f+1)(2I+1)}\delta\omega_{r}.
\end{eqnarray}
It can be noted here that another averaging can be performed. Application of various averaging scheme leads to different values of "central" transition frequency $\nu_{2s-4p}$ lying within the experimental error bars of $3$ kHz. For example, the weighting factor $(2F_r+1)/(2j_r+1)(2I+1)\cdot(2j_f+1)/(2l_f+1)(2s+1)$ instead of Eq.~(\ref{av0}) can be considered, leading to $\nu_{2s-4p} = 616520931628.5$ kHz, which in turn coincides with $\omega_0$ and gives a perfect match for the proton charge radii extracted from the muon hydrogen experiment \cite{HorbHess,Pohl}. Also, the exlusion of the additional factor $1/(2j_f+1)$ in Eq.~(\ref{av0}) leads to the experimental values of $\nu_{1/2}$, $\nu_{3/2}$ and, hence, $\nu_{2s-4p}$ within the error of experiment \cite{Science}.

The results collected in Table~\ref{tab:1} clearly show that "line maximum" can be used to determine the frequency "invariant". First, the nonresonant correction $\delta\omega_r$ can be used as a frequency shift corresponding to the maximum of the line profile. Since $\delta\omega_{r}$ does not dependent on the angle $\theta$, one can simply subtract it from $\omega_{\rm max}$ resulting in $\omega_0$ ("invariant"). Second, the averaging procedure applied here shows a perfect agreement with the experimental results for the frequencies $\nu_{1/2}$ and $\nu_{3/2}$. Thus, there is no contradiction in concepts of transition frequency determination: i) the result reported in \cite{Science} represents the averaged centroid, ii) the same result can be achieved using the partial scattering channels in conjunction with the nonresonant corrections. It is worth noting that the procedure of constructing centroids described above seems to be simpler than the simmetrization (Fano-Voight) procedure employed in \cite{Science}. Subtraction of relatively big numbers (NR corrections, which may reach $1$ MHz \cite{Science}) is always more dangerous than obtaining the same results without subtractions. In centroids, asymmetry cancellation is automatic due to opposite signs of NR corrections for two components.

The situation discussed above, however, can be greatly simplified by choosing a particular decay channel. Based on the experiment \cite{Eikema2001}, it can be asserted that the hyperfine structure of the ground state is resolvable by observing the emission of an excited atom. Thus, the measurement of $2s-4p$-absorption can be restricted to registering the decay channel to the $1s_{1/2}^{F_f=0}$ and/or $1s_{1/2}^ {F_f=1}$ state. These channels are also preferred for another reason (see next section) and can be used to determine the frequency "invariant" as follows. Performing averaging of values listed in Table~\ref{tab:1}, according to Eq. (\ref{3}), one can find
\begin{eqnarray}
\label{1sHFS-a}
\nu_{2s-4p}^{(F_f=0)} = \frac{1}{3}\omega_{\rm max}^{(4p_{1/2})}+\frac{2}{3}\omega_{\rm max}^{(4p_{3/2})}-\Delta_{\rm HFS} 
\nonumber
\\
= 616520931638.7\, {\rm kHz},
\\
\nonumber
\nu_{2s-4p}^{(F_f=1)} = 616520931638.7\, {\rm kHz}.
\end{eqnarray}
Thus, additional averaging over the hyperfine components leads to the same result, which is about $12$ kHz higher than the value reported in \cite{Science}. However, the deviation is much smaller than the corrections $\delta\omega_r$, see the third column in Table~\ref{tab:1}.

Since the corrections $\delta\omega_r$ are theoretical (the analog of such calculations is the fitting of the experimentally observed line profile, see \cite{Jent-NR}), we can use the averaging scheme given by Eq.~(\ref{av0}). This gives
\begin{eqnarray}
\label{1sHFS-b}
\nu_{2s-4p}^{(F_f=0)} = 616520931630.7\, {\rm kHz},
\nonumber
\\
\nu_{2s-4p}^{(F_f=1)} = 616520931627.3\, {\rm kHz},
\\
\nonumber
\nu_{2s-4p}^{(\rm av)} = 616520931628.2\, {\rm kHz}.
\end{eqnarray}
These values are within experimental error with the results \cite{HorbHess,Science} (see also Table~\ref{tab:1}).

Thus, with measured partial channels, the determination of the transition frequency comes down to different possibilities, giving in the end different values, albeit within the experimental error. This is the manifestation of going beyond the resonance approximation. There is no unambiguous definition of the transition frequency, because different averaging schemes can be considered. In the general case, the "central" values of such centroids may vary, although they should be within the experimental error. Nevertheless, the determination of physical constants is very sensitive to this possible difference in values. 
 
So far, we have discussed the problem of determining the transition frequency from the observed line profile when all radiation is detected \cite{Science}. At the same stage one can raise the question of identifying the observed line profile with the absorption process, as in the experiment \cite{Science}. In particular, it was claimed that the found line profile represents absorption, although it was measured by recording the emitted photons. Thus, the authenticity of the separation or the identification of the emission with the absorption process at the level of a few kHz should be carefully checked. This question brings us back to the problem of a detailed description of the process and its corresponding line profile. Within such a description, effects leading to significant asymmetry are possible. In other words, it is worth discussing whether the emission process can affect the absorption profile and vice versa. This is what we will deal with in the next section.

\section{Engaging the emission process in determination the absorption transition frequency}

The fundamental principles governing the detailed description of observed line profile require detailed consideration of all the processes involved in the measurements. For example, the evaluation of nonresonant terms in the photon scattering cross section shows a distinct difference for the cases when all outgoing photons are registered or when a particular scattering process is utilized in determining the transition frequency \cite{Solovyev2020}. The QED formulation of the line profile theory itself and the use of the resonance approximation for the photon scattering cross section require consideration of the process from a stable to a stable state. The exploitation of the metastable state is also admissible. For its part, the determination of the $2s-4p$ absorption frequency is not limited to the two-photon scattering $2s\rightarrow 4p\rightarrow 1s(2s)$ process, but also involves transitions to $3s$ and $3d$ states, which then decay through two-photon emission into the stable $1s$ state (the $2s$ metastable state) \cite{Science}. We drop the discussion about the ambiguity of separating the cascades from the "pure" two-photon emission \cite{LSP-separation}, since the interference between these two types of the probabilities is too small.
 Accordingly, the determination of the $2s-4p$ transition frequency can only be aptly described by taking into account the following cascade processes: $2s+\gamma\rightarrow 4p\rightarrow 3s(3d) +\gamma\rightarrow 2p (3p) + \gamma\rightarrow 1s (2s) + \gamma$. Further, we restrict ourselves to describing the cascade emission to the $1s$ state only but taking into account the hyperfine structure of the levels. 

It is worth noting at once that cascade radiation is also subject to the quantum interference effect when interference between sublevels of hyperfine structure is considered. Within the framework of the above approximations (see also \cite{PhysRevA.92.022514,PhysRevA.92.062506}), a detailed consideration of the $2s+\gamma\rightarrow 4p\rightarrow 3s(3d) +\gamma\rightarrow 2p (3p) + \gamma\rightarrow 1s (2s) + \gamma$ cascade transition should include $4p_{1/2}^{F=1}$ $4p_{3/2}^{F=1}$ states as resonant (denoted below as state $r$), see \cite{Science}, states $3s_{1/2}$ and $3d_{3/2}$, $3d_{5/2}$ as the first cascade state (denoted below as state $a$), and then states $2p_{1/2}$, $2p_{3/2}$, $3p_{1/2}$, $3p_{3/2}$ (denoted below as $b$ state). In our subsequent calculations, we sum over the total atomic momentum $F$ for the states $a$ and $b$. Leaving the resonance term in the cross section (see Appendix~\ref{SuppM1} for details), the effect of QI in the cascade can be described by regarding in the amplitude apart from the resonant states also the nearest by energy states. The amplitude contains now three energy denominators, each of which can be reduced to the absorption resonance denominator using the energy conservation law. Then QIE arise for $a$ and $b$ states exactly as for $r$ states with appropriate widths and energy intervals between resonant and neighboring nonresonant states. 

Considering first the cascade through the $3s$ state, it can be found that the absorption frequency, determined from the extremum condition, is (see the derivation in Appendix~\ref{SuppM1})
\begin{eqnarray}
\label{4}
\omega_{\rm max} = \omega_0 +\delta\omega_r +\delta\omega_a +\delta\omega_b,
\end{eqnarray}
where $\omega_0$ represents $\nu_{1/2}$ or $\nu_{3/2}$, according to the notations in \cite{Science}, 
 $\delta\omega_i$ is given by
\begin{eqnarray}
\label{5}
\delta\omega_i = \frac{f^{(c)}_{\rm nr}}{f^{(c)}_{\rm res}}\frac{\Gamma^2_r}{4\Delta_i}\Upsilon,\qquad\qquad
\\
\nonumber
\Upsilon = \frac{(\Gamma_r+\Gamma_a)^2(\Gamma_a+\Gamma_b)^2}{(\Gamma_r+\Gamma_a)^2(\Gamma_a+\Gamma_b)^2+(\Gamma_r+\Gamma_a)^2\Gamma_r^2+(\Gamma_a+\Gamma_b)^2\Gamma_r^2}.
\end{eqnarray}
Here $i$ is one of states denoted by $r$, $a$, $b$. We define $f^{(c)}_{\rm res}$ as the numerator of the resonance amplitude. In case of the cascade going via the state $3s_{1/2}$ this amplitude corresponds to $2s^{F=0}_{1/2}+\gamma\rightarrow 4p^{F=1}_{1/2(3/2)}\rightarrow 3s_{1/2}+\gamma\rightarrow 2p_{1/2}+\gamma\rightarrow 1s_{1/2}+\gamma$ transition and $f^{(c)}_{\rm nr}$ corresponds to one of nonresonant contributions $2s_{1/2}+\gamma\rightarrow 4p^{F=1}_{1/2(3/2)} \rightarrow a +\gamma\rightarrow b+\gamma\rightarrow 1s_{1/2}+\gamma$, $\Delta_i$ represents the energy splitting between the states $r$, $a$ or $b$ (as, for example, $\Delta_r\equiv E_{4p^{F=1}_{3/2}}-E_{4p^{F=1}_{1/2}}$), and $\delta\omega_i$ correction is written out in the lowest order.

Assuming that all outgoing photons are directed to the same detector (i.e. photon direction vectors, $\vec{\nu}$, are aligned, $\vec{\nu}_{4p-3s}\parallel\vec{\nu}_{3s-2p(3p)}\parallel\vec{\nu}_{2p-1s}$), the NR corrections are
\begin{eqnarray}
\label{6}
\delta\omega_r = -\frac{1}{2}\left(1-3\cos 2\theta\right)\frac{\Gamma^2_r}{4\Delta_r}\Upsilon,
\\
\nonumber
\delta\omega_a = 0,
\\
\nonumber
\delta\omega_b = 2\frac{\Gamma^2_r}{4\Delta_b}\Upsilon.
\end{eqnarray}
Here and below $\theta$ denotes the angle between the polarization vector of the incident photon and the direction vector of the emitted photons (this corresponds to the conditions of experiment \cite{Science}), and the numerical factors arise from the ratio of the radial parts of the amplitudes. It may be noted that when $F_f$ is fixed, the results \cite{Solovyev2020} are reconstructed. Thus, the angular dependence in Eq.~(\ref{6}) is due to summation over the total atomic momenta in the amplitudes $f^{(c)}_{\rm res}$ and $f^{(c)}_{\rm nr}$.

It is found that the correction $\delta\omega_a$ is equal to zero. This is valid only for the case when photons are registered in one direction, otherwise the correction does not vanish at all. Using the values $\Gamma_r=1.2941\times 10^7$ Hz, $\Delta_r = 1\,367\,433.3$ kHz
, $\Gamma_{3d}=1.0295\times 10^7$ Hz, $\Gamma_{3p}=3.0208\times 10^6$ Hz, $\Gamma_{3s}=1.0054\times 10^6$ Hz, $\Gamma_{2p}=9.97624\times 10^7$ Hz, $\Delta_{3p} =3\,241\,327.3$ kHz, $\Delta_{2p} =10\,939\,469.7$ kHz \cite{HorbHess}, the frequency shifts are given in Table~\ref{tab:2} at various angles $\theta$.
\begin{widetext}
\begin{center}
\begin{table}[ht!]
\renewcommand{\arraystretch}{1.2}
\caption{Numerical values of nonresonant shifts and total contribution $\delta\omega_\Sigma$, multiplied by a branching ratio factor of $W_{4p-n_al_a}/\Gamma_{4p}$. The value of the angle at which the maximum ($\theta_{\rm max}$), minimum ($\theta_{\rm min}$) values are reached is given, as well as the values of NR corrections at the magic angle ($\theta_{\rm m}$). The values of the angle at which the total contribution is zero (if it exists), $\theta_0$, cascade fraction (branching ratio) and the process in question, are given in the detached row. 
 All values are given in kHz.}
\label{tab:2}
\begin{tabular*}{\textwidth}{@{\extracolsep{\fill}}l c c c}
\hline
Angle  & $\delta\omega_r+\delta\omega_{3p}$ in kHz & $\delta\omega_r+\delta\omega_{2p}$ in kHz & $\delta\omega_{\Sigma}$ in kHz\\
\hline
\hline
\multicolumn{4}{c}{$2s_{1/2}^{F=0}\rightarrow 4p^{F=1}_{1/2} \rightarrow 3s_{1/2}\rightarrow 2p_{1/2} \rightarrow 1s_{1/2}$; $\theta_0=\pm0.403619$;}\\
\multicolumn{4}{c}{$2s_{1/2}^{F=0}\rightarrow 4p^{F=1}_{1/2} \rightarrow 3s_{1/2}\rightarrow 3p_{1/2} \rightarrow 1s_{1/2}$;  $W_{4p-3s}/\Gamma_{4p}\approx 0.0377$} \\

\hline
\hline
 $\theta_{\rm min}=0$ & $-3.388$ & $-17.342$ & $-0.781$\\
 
 $\theta_{\rm max}=\pi/2$ & $61.587$ & $52.008$ & $4.282$\\
 
 $\theta_{\rm m}$ & $39.942$  & $28.898$ & $2.595$\\
\hline
\hline

\multicolumn{4}{c}{$2s_{1/2}^{F=0}\rightarrow 4p^{F=1}_{1/2} \rightarrow 3d_{3/2}\rightarrow 2p_{1/2} \rightarrow 1s_{1/2}$; $\theta_0=\pm 0.403605$;}\\
\multicolumn{4}{c}{$2s_{1/2}^{F=0}\rightarrow 4p^{F=1}_{1/2} \rightarrow 3d_{3/2}\rightarrow 3p_{3/2} \rightarrow 1s_{1/2}$;  $W_{4p-3s}/\Gamma_{4p}\approx 0.0043$} \\

\hline
\hline
 $\theta_{\rm min}=0$ & $-0.339$ & $-1.734$ & $-9.\times 10^{-3}$\\
 
 $\theta_{\rm max}=\pi/2$ & $6.164$ & $5.203$ & $4.9\times 10^{-2}$\\
 
 $\theta_{\rm m}$ & $3.997$  & $2.890$ & $3.0\times 10^{-2}$\\
\hline
\hline

\multicolumn{4}{c}{$2s_{1/2}^{F=0}\rightarrow 4p^{F=1}_{1/2} \rightarrow 3s_{1/2}\rightarrow 2p_{3/2} \rightarrow 1s_{1/2}$\footnotemark{}\strut; $\theta_0=\pm 0.637414$; $W_{4p-3d}/\Gamma_{4p}\approx 0.0377$}\\

\hline
\hline
 $\theta_{\rm min}=0$ & $--$ & $-21.679$ & $-0.817$\\
 
 $\theta_{\rm max}=\pi/2$ & $--$ & $43.343$ & $1.634$\\
 
 $\theta_{\rm m}$ & $--$ & $21.675$ & $0.817$\\
\hline
\hline

\multicolumn{4}{c}{$2s_{1/2}^{F=0}\rightarrow 4p^{F=1}_{1/2} \rightarrow 3d_{3/2}\rightarrow 2p_{3/2} \rightarrow 1s_{1/2}$;}\\
\multicolumn{4}{c}{$2s_{1/2}^{F=0}\rightarrow 4p^{F=1}_{1/2} \rightarrow 3d_{3/2}\rightarrow 3p_{3/2} \rightarrow 1s_{1/2}$;  $W_{4p-3d}/\Gamma_{4p}\approx 0.0043$} \\

\hline
\hline
 $\theta_{\rm max}=0$ & $-224.435$ & $-74.460$ & $-1.285$\\
 
 $\theta_{\rm min}=\pi/2$ & $-218.192$ & $-67.534$ & $-1.229$\\
 
 $\theta_{\rm m}$ & $-220.274$ & $-69.843$ & $-1.247$\\

\hline
\hline
\end{tabular*}
\end{table}
\footnotetext{There is no decay to the $3p_{3/2}$ state}
\end{center}
\end{widetext}

As it follows from Eq.~(\ref{6}), the correction $\delta\omega_r$ is zero at angles $\theta=\pm 1/2 \arccos(1/3)+\pi k$ ($k$ is an integer), other than the magic angle $\theta_{\rm m} = \arccos (1/\sqrt{3})$. Solving the equation for the total correction $\delta\omega_{\Sigma}(\theta_0) =\delta\omega_r+\delta\omega_{2p} =0$, one can find the angle at which it vanishes (if it exists, see Fig.~\ref{fig2s} in Appendix~\ref{me}). According to \cite{Science}, the fraction of the cascade process is about $4\%$ of all photons captured by detector. It can be found as a ratio of partial transition probability to the level width: $W_{4p-3s}/\Gamma_{4p}\approx 0.0377$, $W_{4p-3d}/\Gamma_{4p}\approx 0.0043$, which were used to obtain the total contribution, $\delta\omega_{\Sigma}$. It should be noted that using these coefficients to obtain the transition frequency $\nu_{2s-4p}$ instead of the rms value (see Table~\ref{tab:1}) leads to a different centroid. 

Repeating the calculations for the cascade transition going through the state $3d_{3/2}$ for the $\nu_{1/2}$ frequency, one can obtain
\begin{eqnarray}
\label{7}
\delta\omega_r = \frac{1}{20}\left(1-3\cos 2\theta\right)\frac{\Gamma^2_r}{4\Delta_r}\Upsilon,
\\
\nonumber
\delta\omega_a = 0,
\\
\nonumber
\delta\omega_{b} = \frac{1}{5}\frac{\Gamma^2_r}{4\Delta_b}\Upsilon.
\end{eqnarray}
Here the resonance amplitude corresponds to the $2s^{F=0}_{1/2}+\gamma\rightarrow 4p^{F=1}_{1/2}\rightarrow 3d_{3/2}+\gamma\rightarrow 2p_{1/2}+\gamma\rightarrow 1s_{1/2}+\gamma$ and the nonresonant amplitudes are related to i) $2s^{F=0}_{1/2}+\gamma\rightarrow 4p^{F=1}_{3/2}\rightarrow 3d_{3/2}+\gamma\rightarrow 2p_{1/2}+\gamma\rightarrow 1s_{1/2}+\gamma$, $\Delta_r\equiv E_{4p^{F=1}_{3/2}}-E_{4p^{F=1}_{1/2}}$ and ii) $2s^{F=0}_{1/2}+\gamma\rightarrow 4p^{F=1}_{1/2}\rightarrow 3d_{3/2}+\gamma\rightarrow 2p_{3/2}+\gamma\rightarrow 1s_{1/2}+\gamma$, $\Delta_b\equiv E_{2p_{3/2}}-E_{2p_{1/2}}, E_{3p_{3/2}}-E_{3p_{1/2}}$ decay channels. The numerical values, multiplied by a factor of $0.0043$ according to the contribution of the cascade to the transition frequency measurements in \cite{Science}, are collected in the second segment of Table~\ref{tab:2}. When the resonant channel is treated as passing through the $2p_{3/2}(3p_{3/2})$ state, the numerical results are presented in the third and fourth segments of Table~\ref{tab:2} (the corresponding graphs are illustrated Figs.~\ref{fig2s}-\ref{fig5s} in Appendix~\ref{SuppM1}).

From the above analysis, one can conclude that the contribution of QIEc is in general significant, but is suppressed by the relative fraction of the cascade process in the total radiation recorded in the experiment \cite{Science}. Nevertheless, the interfering paths in the cascade process are influential at the level of several kilohertz on the absorption line profile. This asymmetry can be expressed through a nonresonant correction to the transition frequency, defined here as the maximum of the line profile. This corrections does not vanish at the magic angle, see Table~\ref{tab:2}. In practice, this means that the symmetrization procedure applied in \cite{Science} has reduced the QIEc to the value estimated here at the magic angle because an appropriate asymmetry parameter was used. Although the value of $\delta\omega_\Sigma$ is within the error bars of the experiment, for the "central" value with the same uncertainty a frequency shift at the kHz level can be expected.

Above we gave as a demonstration an analysis on measuring the transition frequency $\nu_{1/2}$. Similarly, one can perform calculations for the transition frequency $\nu_{3/2}$. Discarding for brevity the details of the calculations, the numerical results are presented in Table~\ref{tab:3}.
\begin{widetext}
\begin{center}
\begin{table}[h!]
\renewcommand{\arraystretch}{1.2}
\caption{Numerical values of nonresonant shifts for specified cascade transitions corresponding to the frequency $\nu_{3/2}= E_{4p_{3/2}^{F=1}} - E_{2s_{1/2}^{F=0}}$ and total contribution $\delta\omega_\Sigma$, multiplied by a factor of $W_{4p-nl}/\Gamma_{4p}$. The notations are the same as in Table~\ref{tab:2}. All values are given in Hz.}
\label{tab:3}
\begin{tabular*}{\textwidth}{@{\extracolsep{\fill}} l c c c}

\hline
Angle & $\delta\omega_r+\omega_{3p}$ in Hz & $\delta\omega_r+\omega_{2p}$ in Hz & $\omega_{\Sigma}$ in Hz\\

\hline
\hline
\multicolumn{4}{c}{$2s_{1/2}^{F=0}\rightarrow 4p^{F=1}_{3/2} \rightarrow 3s_{1/2}\rightarrow 2p_{1/2} \rightarrow 1s_{1/2}$;}\\
\multicolumn{4}{c}{$2s_{1/2}^{F=0}\rightarrow 4p^{F=1}_{3/2} \rightarrow 3s_{1/2}\rightarrow 3p_{1/2} \rightarrow 1s_{1/2}$;  $W_{4p-3s}/\Gamma_{4p}\approx 0.0377$} \\
\hline
 $\theta_{\rm max}=0$ & $39.954$ & $28.902$ & $2.596$\\
 
 $\theta_{\rm min}=\pi/2$ & $7.446$ & $-5.780$ & $6.3\times 10^{-2}$\\
 
 $\theta_{\rm m}$ & $11.058$ & $-1.927$ & $0.344$\\
\hline
\hline

\multicolumn{4}{c}{$2s_{1/2}^{F=0}\rightarrow 4p^{F=1}_{3/2} \rightarrow 3s_{1/2}\rightarrow 2p_{3/2} \rightarrow 1s_{1/2}$\footnote{}; $\theta_0=0.528655$;}\\

\hline
 $\theta_{\rm max}=0$ & $--$ & $21.679$ & $0.817$\\
 
 $\theta_{\rm min}=\pi/2$ & $--$ & $-13.006$ & $-0.490$\\
 
 $\theta_{\rm m}$ & $--$ & $-9.152$ & $-0.345$\\
\hline
\hline

\multicolumn{4}{c}{$2s_{1/2}^{F=0}\rightarrow 4p^{F=1}_{3/2} \rightarrow 3d_{3/2}\rightarrow 2p_{1/2} \rightarrow 1s_{1/2}$; $\theta_0=0.651478$;}\\
\multicolumn{4}{c}{$2s_{1/2}^{F=0}\rightarrow 4p^{F=1}_{3/2} \rightarrow 3d_{3/2}\rightarrow 3p_{1/2} \rightarrow 1s_{1/2}$;  $W_{4p-3s}/\Gamma_{4p}\approx 0.0043$} \\

\hline
 $\theta_{\rm max}=0$ & $9.571$ & $8.836$ & $7.9\times 10^{-2}$\\
 
 $\theta_{\rm min}=\pi/2$ & $-106.438$ & $-114.926$ & $-0.952$\\
 
 $\theta_{\rm m}$ & $-16.235$ & $-18.690$ & $-0.150$\\
\hline
\hline

\multicolumn{4}{c}{$2s_{1/2}^{F=0}\rightarrow 4p^{F=1}_{3/2} \rightarrow 3d_{3/2}\rightarrow 2p_{3/2} \rightarrow 1s_{1/2}$;}\\
\multicolumn{4}{c}{$2s_{1/2}^{F=0}\rightarrow 4p^{F=1}_{3/2} \rightarrow 3d_{3/2}\rightarrow 3p_{3/2} \rightarrow 1s_{1/2}$;  $W_{4p-3s}/\Gamma_{4p}\approx 0.0043$} \\

\hline
 $\theta_{\rm max}=0$ & $-37.927$ & $-6.192$ & $-0.190$\\
 
 $\theta_{\rm min}=\pi/2$ & $-153.437$ & $-129.889$ & $-1.218$\\
 
 $\theta_{\rm m}$ & $-63.665$ & $-33.709$ & $-0.419$\\

\hline
\hline
\end{tabular*}
\end{table}
\footnotetext{There is no decay to the $3p_{3/2}$ state}
\end{center}
\end{widetext}

The main conclusion of this section, however, which follows from the QIEc analysis, is that the Fano profile obtained with the cascade process should be involved in the symmetrization procedure \cite{Science}. Then, for example, several parameters related to the asymmetry of the line profile due to different processes should be used to best fit the experimental data. The asymmetry parameters do not necessarily depend equally on the angle and can in principle be treated as independent of each other. In fact, we can state that modern spectroscopic experiments represent a frontier leading to the next generation of experiments in which the problem of resonance approximation will have a decisive role. 

As a consequence, consideration of the cascade process affecting the determination of the absorption transition frequency shows the inseparability of the absorption and emission processes in describing the line profile beyond the resonance approximation. The "central" value is expected to be shifted at the kHz level. Gathering together the results of this study it can be seen that the photon scattering process used in the experiment \cite{Science} is rather complicated and has to include with necessity the analysis of the observed line profile asymmetry. This asymmetry is caused not only by the effect of quantum interference for resonant absorption, but also by cascading emission processes. Assuming the need to increase the experimental accuracy (e.g., for precision determination of physical constants), the analysis of cascade processes will be increasingly required for experiments of the type \cite{Science} (when all radiation is detected). However, for the measured $2s-4p$ line in the \cite{Science} experiment it is possible to distinguish a case unaffected by QIEc. The main contribution to the emission comes from the $4p-1s$ emission line, in which there is no cascade. Consequently, this scattering channel is preferable for determining the transition frequency. An appropriate experiment can be performed by registering emitted photons with a certain energy equal to $4p-1s$. Such experiments should be more accurate.

\section{Conclusions}
In this paper, the principles of transition frequency determination based on precision spectroscopic experiments and, as a consequence, the accuracy of its definition are discussed in detail. The assumption that the asymmetry of the line profile caused by nonresonant terms in the scattering cross section limits the accuracy of the transition frequency determination, see, e.g., \cite{2001,Jent-NR}, has recently been countered in \cite{Science}, where the value of transition frequency was obtained with accuracy far beyond the NR corrections. This seeming controversy follows from the new situation arising in atomic resonant spectroscopy. The main goal of \cite{Science} was to obtain the empirical frequency value which may be used for extracting the physical constants from the experimental data. For this purpose, it is necessary to extract from this data an artificial symmetric line profile. The central (as well as maximum) frequency value in this contour can be directly compared with the theoretical energy difference of atomic levels, for which the asymmetry subtraction procedure (Fano-Voghit) was applied in \cite{Science}.

In \cite{2001,Jent-NR,Solovyev2020} the real asymmetric line profile distorted by measurement was considered. The natural way to define the transition frequency value for such a line shape is to choose the maximum, i.e. most probable value. However, then the result (transition frequency value) begins to depent on the decay channel for the excited state \cite{Solovyev2020}. This happens only beyond the resonance approximation and the obtained transition frequencies differ by NR corrections. This led to the conclusion that beyond the resonance approximation it is not possible to define uniquelly the atomic transition frequency.

In this paper it was demonstrated that the extraction of the symmetric line profile and determination of its frequency can be succesfully made for any process with fixed decay channel with the same result (the frequency invariant or the frequency standard) as in \cite{Science}. Definition of the frequency standard was not the main goal of \cite{Science}, but actually the frequency "invariant" obtained with many digits beyond the NR correction value may serve as such standard.

The transition frequency value obtained as a result of processing the data from the experiment \cite{Science} pertains rather to a special case and is, according to the presented analysis, the first step in measuring the transition frequency beyond the resonance approximation. Although increasing the accuracy of measurements by an order of magnitude, this type of experiment still faces obstacles related to i) defining the transition frequency; ii) the accuracy of the subtraction procedure and its replicability; iii) the need to account for asymmetry due to cascade emission. All these circumstances are precisely the result of going beyond the resonance approximation.

In particular, we found that the "line maximum" of the observed profile can be used as a transition frequency with the same rights as the "line center". First, the transition frequency values, $\omega_0$ (tabulated in \cite{HorbHess}), can be found by subtracting the corresponding NR correction from $\omega_{\rm max}$. This implies promptly from the definition of the maximum of the line profile, Eq.~(\ref{1}). Second, different processing of the measured $\omega_{\rm max}$ can be performed, resulting in deviations of the transition frequency values within the experimental error, but with a different "central" value. One such scheme leads exactly to the values specified in \cite{Science}, which can be seen as rectifying discrepancies in the definitions of the transition frequency beyond the resonance approximation. Third, because of the inseparability of the absorption process from the emission, the asymmetry caused by the cascade radiation leads to an additional shift of the absorption transition frequency. Despite its fractional contribution at the $4\%$ level, the corresponding distortion of the line profile remains significant for experiments of the type \cite{Science}. This circumstance leads to a more preferable partial photon scattering channel for measurements: $2s-4p-1s$, which is easily achieved by matching with $4p-1s$ emitted photons.

The presence of cascade emission in the measurement of the $2s-4p$ transition frequency is the crucial factor in the context of the experiment \cite{Science}. The invisible to the naked eye corresponding asymmetry 
 has to be taken into account when fitting the observed line profile. The Fano-Voigt contour used in \cite{Science} should be modified at least in the linear approximation, see the procedure in \cite{Jent-NR}, using the formulas for the photon scattering cross section with the presence of a cascade emission process given in Appendix~\ref{SuppM1}. As a consequence, the resulting profile should depend on the parameters of the intermediate atomic states involved in the cascade emission, such as level width and energy (energy splitting).

The issue of reconciliation among the concepts of optical frequency standards can be considered in accordance with the following stipulations. i) Unique definition of transition frequency can be done only within the resonant approximation, going beyond it requires further theoretical processing. Extending the definition of the transition frequency beyond the resonance approximation necessarily involves a detailed description of the nonresonant effects leading to the asymmetry of the observed line profile. ii) With NR corrections transition frequency begins to depend on the preparation of the initial state, on the decay to the final state and on the geometry of experiment. iii) Dependence on the lowest order NR corrections can be eliminated from the experimental data by using "magic angle" or by direct subtraction of asymmetric contributions from these data \cite{Science}. In this way the transition frequency value invariant with respect to the geometry of experiment ("line center" according to \cite{Science}) can be obtained. iv) The determination of the "invariant" transition frequency, for the absorption process, has to carefully consider the emission and vice versa. v) Still, the "line maximum" can be used to determine the transition frequency "invariant". For the same obstacles caused by nonresonant effects, the invariant can be found using the averaging procedure or subtracting the NR corrections. vi) The strategy opposite to the experiment \cite{Science} is preferable. The detection of emitted photons corresponding to decay to the ground state eliminates the asymmetry of the line profile caused by radiation.

Throughout the paper, we explore "line maximum" instead of "line center", what corresponds to the choice of most probable value. We also include in the consideration all possible decay channels as the equivalent sources for determination of the resonant transition frequency. In this way we obtain, in case of N channels, N equivalent (not equal) resonant transition frequencies differing from each other by NR corrections. All these values for resonant transition frequency are automatically invariant with respect to the geometry of experiment, neither "magic angle", nor asymmetry subtraction from experimental data are required. The "line center" value for the resonant transition frequency as well as all "line maximum" values are equally suitable for the choice as frequency standards in conjunction with appropriate evaluation of NR corrections. The question is which conditions are easier to reproduce in every laboratory: use "magic angle", asymmetry subtraction procedure or to choose the particular decay channel to the final state. The answer probably is different for the different candidates to the frequency standard.

\section*{Acknowledgments}
This work was supported by the Russian Science Foundation under grant No. 22-12-00043.

\bibliography{NR-SZAL}

\begin{thebibliography}{41}%
\makeatletter
\providecommand \@ifxundefined [1]{%
 \@ifx{#1\undefined}
}%
\providecommand \@ifnum [1]{%
 \ifnum #1\expandafter \@firstoftwo
 \else \expandafter \@secondoftwo
 \fi
}%
\providecommand \@ifx [1]{%
 \ifx #1\expandafter \@firstoftwo
 \else \expandafter \@secondoftwo
 \fi
}%
\providecommand \natexlab [1]{#1}%
\providecommand \enquote  [1]{``#1''}%
\providecommand \bibnamefont  [1]{#1}%
\providecommand \bibfnamefont [1]{#1}%
\providecommand \citenamefont [1]{#1}%
\providecommand \href@noop [0]{\@secondoftwo}%
\providecommand \href [0]{\begingroup \@sanitize@url \@href}%
\providecommand \@href[1]{\@@startlink{#1}\@@href}%
\providecommand \@@href[1]{\endgroup#1\@@endlink}%
\providecommand \@sanitize@url [0]{\catcode `\\12\catcode `\$12\catcode
  `\&12\catcode `\#12\catcode `\^12\catcode `\_12\catcode `\%12\relax}%
\providecommand \@@startlink[1]{}%
\providecommand \@@endlink[0]{}%
\providecommand \url  [0]{\begingroup\@sanitize@url \@url }%
\providecommand \@url [1]{\endgroup\@href {#1}{\urlprefix }}%
\providecommand \urlprefix  [0]{URL }%
\providecommand \Eprint [0]{\href }%
\providecommand \doibase [0]{http://dx.doi.org/}%
\providecommand \selectlanguage [0]{\@gobble}%
\providecommand \bibinfo  [0]{\@secondoftwo}%
\providecommand \bibfield  [0]{\@secondoftwo}%
\providecommand \translation [1]{[#1]}%
\providecommand \BibitemOpen [0]{}%
\providecommand \bibitemStop [0]{}%
\providecommand \bibitemNoStop [0]{.\EOS\space}%
\providecommand \EOS [0]{\spacefactor3000\relax}%
\providecommand \BibitemShut  [1]{\csname bibitem#1\endcsname}%
\let\auto@bib@innerbib\@empty
\bibitem [{\citenamefont {Weisskopf}\ and\ \citenamefont
  {Wigner}(1930)}]{Weisskopf1930}%
  \BibitemOpen
  \bibfield  {author} {\bibinfo {author} {\bibfnamefont {V.}~\bibnamefont
  {Weisskopf}}\ and\ \bibinfo {author} {\bibfnamefont {E.}~\bibnamefont
  {Wigner}},\ }\href {\doibase 10.1007/BF01336768} {\bibfield  {journal}
  {\bibinfo  {journal} {Zeitschrift f{\"u}r Physik}\ }\textbf {\bibinfo
  {volume} {63}},\ \bibinfo {pages} {54} (\bibinfo {year} {1930})}\BibitemShut
  {NoStop}%
\bibitem [{\citenamefont {Low}(1952)}]{Low}%
  \BibitemOpen
  \bibfield  {author} {\bibinfo {author} {\bibfnamefont {F.}~\bibnamefont
  {Low}},\ }\href {\doibase 10.1103/PhysRev.88.53} {\bibfield  {journal}
  {\bibinfo  {journal} {Phys. Rev.}\ }\textbf {\bibinfo {volume} {88}},\
  \bibinfo {pages} {53} (\bibinfo {year} {1952})}\BibitemShut {NoStop}%
\bibitem [{\citenamefont {Andreev}\ \emph {et~al.}(2008)\citenamefont
  {Andreev}, \citenamefont {Labzowsky}, \citenamefont {Plunien},\ and\
  \citenamefont {Solovyev}}]{ANDREEV2008135}%
  \BibitemOpen
  \bibfield  {author} {\bibinfo {author} {\bibfnamefont {O.~Y.}\ \bibnamefont
  {Andreev}}, \bibinfo {author} {\bibfnamefont {L.~N.}\ \bibnamefont
  {Labzowsky}}, \bibinfo {author} {\bibfnamefont {G.}~\bibnamefont {Plunien}},
  \ and\ \bibinfo {author} {\bibfnamefont {D.~A.}\ \bibnamefont {Solovyev}},\
  }\href {\doibase https://doi.org/10.1016/j.physrep.2007.10.003} {\bibfield
  {journal} {\bibinfo  {journal} {Physics Reports}\ }\textbf {\bibinfo {volume}
  {455}},\ \bibinfo {pages} {135} (\bibinfo {year} {2008})}\BibitemShut
  {NoStop}%
\bibitem [{\citenamefont {Zalialiutdinov}\ \emph {et~al.}(2018)\citenamefont
  {Zalialiutdinov}, \citenamefont {Solovyev}, \citenamefont {Labzowsky},\ and\
  \citenamefont {Plunien}}]{ZALIALIUTDINOV20181}%
  \BibitemOpen
  \bibfield  {author} {\bibinfo {author} {\bibfnamefont {T.~A.}\ \bibnamefont
  {Zalialiutdinov}}, \bibinfo {author} {\bibfnamefont {D.~A.}\ \bibnamefont
  {Solovyev}}, \bibinfo {author} {\bibfnamefont {L.~N.}\ \bibnamefont
  {Labzowsky}}, \ and\ \bibinfo {author} {\bibfnamefont {G.}~\bibnamefont
  {Plunien}},\ }\href {\doibase https://doi.org/10.1016/j.physrep.2018.02.003}
  {\bibfield  {journal} {\bibinfo  {journal} {Physics Reports}\ }\textbf
  {\bibinfo {volume} {737}},\ \bibinfo {pages} {1} (\bibinfo {year} {2018})},\
  \bibinfo {note} {qED theory of multiphoton transitions in atoms and
  ions}\BibitemShut {NoStop}%
\bibitem [{\citenamefont {Niering}\ \emph {et~al.}(2000)\citenamefont
  {Niering}, \citenamefont {Holzwarth}, \citenamefont {Reichert}, \citenamefont
  {Pokasov}, \citenamefont {Udem}, \citenamefont {Weitz}, \citenamefont
  {H\"ansch}, \citenamefont {Lemonde}, \citenamefont {Santarelli},
  \citenamefont {Abgrall}, \citenamefont {Laurent}, \citenamefont {Salomon},\
  and\ \citenamefont {Clairon}}]{PhysRevLett.84.5496}%
  \BibitemOpen
  \bibfield  {author} {\bibinfo {author} {\bibfnamefont {M.}~\bibnamefont
  {Niering}}, \bibinfo {author} {\bibfnamefont {R.}~\bibnamefont {Holzwarth}},
  \bibinfo {author} {\bibfnamefont {J.}~\bibnamefont {Reichert}}, \bibinfo
  {author} {\bibfnamefont {P.}~\bibnamefont {Pokasov}}, \bibinfo {author}
  {\bibfnamefont {T.}~\bibnamefont {Udem}}, \bibinfo {author} {\bibfnamefont
  {M.}~\bibnamefont {Weitz}}, \bibinfo {author} {\bibfnamefont {T.~W.}\
  \bibnamefont {H\"ansch}}, \bibinfo {author} {\bibfnamefont {P.}~\bibnamefont
  {Lemonde}}, \bibinfo {author} {\bibfnamefont {G.}~\bibnamefont {Santarelli}},
  \bibinfo {author} {\bibfnamefont {M.}~\bibnamefont {Abgrall}}, \bibinfo
  {author} {\bibfnamefont {P.}~\bibnamefont {Laurent}}, \bibinfo {author}
  {\bibfnamefont {C.}~\bibnamefont {Salomon}}, \ and\ \bibinfo {author}
  {\bibfnamefont {A.}~\bibnamefont {Clairon}},\ }\href {\doibase
  10.1103/PhysRevLett.84.5496} {\bibfield  {journal} {\bibinfo  {journal}
  {Phys. Rev. Lett.}\ }\textbf {\bibinfo {volume} {84}},\ \bibinfo {pages}
  {5496} (\bibinfo {year} {2000})}\BibitemShut {NoStop}%
\bibitem [{\citenamefont {Matveev}\ \emph {et~al.}(2013)\citenamefont
  {Matveev}, \citenamefont {Parthey}, \citenamefont {Predehl}, \citenamefont
  {Alnis}, \citenamefont {Beyer}, \citenamefont {Holzwarth}, \citenamefont
  {Udem}, \citenamefont {Wilken}, \citenamefont {Kolachevsky}, \citenamefont
  {Abgrall}, \citenamefont {Rovera}, \citenamefont {Salomon}, \citenamefont
  {Laurent}, \citenamefont {Grosche}, \citenamefont {Terra}, \citenamefont
  {Legero}, \citenamefont {Schnatz}, \citenamefont {Weyers}, \citenamefont
  {Altschul},\ and\ \citenamefont {H\"ansch}}]{PhysRevLett.110.230801}%
  \BibitemOpen
  \bibfield  {author} {\bibinfo {author} {\bibfnamefont {A.}~\bibnamefont
  {Matveev}}, \bibinfo {author} {\bibfnamefont {C.~G.}\ \bibnamefont
  {Parthey}}, \bibinfo {author} {\bibfnamefont {K.}~\bibnamefont {Predehl}},
  \bibinfo {author} {\bibfnamefont {J.}~\bibnamefont {Alnis}}, \bibinfo
  {author} {\bibfnamefont {A.}~\bibnamefont {Beyer}}, \bibinfo {author}
  {\bibfnamefont {R.}~\bibnamefont {Holzwarth}}, \bibinfo {author}
  {\bibfnamefont {T.}~\bibnamefont {Udem}}, \bibinfo {author} {\bibfnamefont
  {T.}~\bibnamefont {Wilken}}, \bibinfo {author} {\bibfnamefont
  {N.}~\bibnamefont {Kolachevsky}}, \bibinfo {author} {\bibfnamefont
  {M.}~\bibnamefont {Abgrall}}, \bibinfo {author} {\bibfnamefont
  {D.}~\bibnamefont {Rovera}}, \bibinfo {author} {\bibfnamefont
  {C.}~\bibnamefont {Salomon}}, \bibinfo {author} {\bibfnamefont
  {P.}~\bibnamefont {Laurent}}, \bibinfo {author} {\bibfnamefont
  {G.}~\bibnamefont {Grosche}}, \bibinfo {author} {\bibfnamefont
  {O.}~\bibnamefont {Terra}}, \bibinfo {author} {\bibfnamefont
  {T.}~\bibnamefont {Legero}}, \bibinfo {author} {\bibfnamefont
  {H.}~\bibnamefont {Schnatz}}, \bibinfo {author} {\bibfnamefont
  {S.}~\bibnamefont {Weyers}}, \bibinfo {author} {\bibfnamefont
  {B.}~\bibnamefont {Altschul}}, \ and\ \bibinfo {author} {\bibfnamefont
  {T.~W.}\ \bibnamefont {H\"ansch}},\ }\href {\doibase
  10.1103/PhysRevLett.110.230801} {\bibfield  {journal} {\bibinfo  {journal}
  {Phys. Rev. Lett.}\ }\textbf {\bibinfo {volume} {110}},\ \bibinfo {pages}
  {230801} (\bibinfo {year} {2013})}\BibitemShut {NoStop}%
\bibitem [{\citenamefont {Labzowsky}\ \emph {et~al.}(2001)\citenamefont
  {Labzowsky}, \citenamefont {Solovyev}, \citenamefont {Plunien},\ and\
  \citenamefont {Soff}}]{2001}%
  \BibitemOpen
  \bibfield  {author} {\bibinfo {author} {\bibfnamefont {L.~N.}\ \bibnamefont
  {Labzowsky}}, \bibinfo {author} {\bibfnamefont {D.~A.}\ \bibnamefont
  {Solovyev}}, \bibinfo {author} {\bibfnamefont {G.}~\bibnamefont {Plunien}}, \
  and\ \bibinfo {author} {\bibfnamefont {G.}~\bibnamefont {Soff}},\ }\href
  {\doibase 10.1103/PhysRevLett.87.143003} {\bibfield  {journal} {\bibinfo
  {journal} {Phys. Rev. Lett.}\ }\textbf {\bibinfo {volume} {87}},\ \bibinfo
  {pages} {143003} (\bibinfo {year} {2001})}\BibitemShut {NoStop}%
\bibitem [{\citenamefont {Jentschura}\ and\ \citenamefont
  {Mohr}(2002)}]{Jent-NR}%
  \BibitemOpen
  \bibfield  {author} {\bibinfo {author} {\bibfnamefont {U.~D.}\ \bibnamefont
  {Jentschura}}\ and\ \bibinfo {author} {\bibfnamefont {P.~J.}\ \bibnamefont
  {Mohr}},\ }\href {\doibase 10.1139/p02-019} {\bibfield  {journal} {\bibinfo
  {journal} {Canadian Journal of Physics}\ }\textbf {\bibinfo {volume} {80}},\
  \bibinfo {pages} {633} (\bibinfo {year} {2002})},\ \Eprint
  {http://arxiv.org/abs/https://doi.org/10.1139/p02-019}
  {https://doi.org/10.1139/p02-019} \BibitemShut {NoStop}%
\bibitem [{\citenamefont {Beyer}\ \emph {et~al.}(2017)\citenamefont {Beyer},
  \citenamefont {Maisenbacher}, \citenamefont {Matveev}, \citenamefont {Pohl},
  \citenamefont {Khabarova}, \citenamefont {Grinin}, \citenamefont {Lamour},
  \citenamefont {Yost}, \citenamefont {H\"{a}nsch}, \citenamefont
  {Kolachevsky},\ and\ \citenamefont {Udem}}]{Science}%
  \BibitemOpen
  \bibfield  {author} {\bibinfo {author} {\bibfnamefont {A.}~\bibnamefont
  {Beyer}}, \bibinfo {author} {\bibfnamefont {L.}~\bibnamefont {Maisenbacher}},
  \bibinfo {author} {\bibfnamefont {A.}~\bibnamefont {Matveev}}, \bibinfo
  {author} {\bibfnamefont {R.}~\bibnamefont {Pohl}}, \bibinfo {author}
  {\bibfnamefont {K.}~\bibnamefont {Khabarova}}, \bibinfo {author}
  {\bibfnamefont {A.}~\bibnamefont {Grinin}}, \bibinfo {author} {\bibfnamefont
  {T.}~\bibnamefont {Lamour}}, \bibinfo {author} {\bibfnamefont {D.~C.}\
  \bibnamefont {Yost}}, \bibinfo {author} {\bibfnamefont {T.~W.}\ \bibnamefont
  {H\"{a}nsch}}, \bibinfo {author} {\bibfnamefont {N.}~\bibnamefont
  {Kolachevsky}}, \ and\ \bibinfo {author} {\bibfnamefont {T.}~\bibnamefont
  {Udem}},\ }\href {\doibase 10.1126/science.aah6677} {\bibfield  {journal}
  {\bibinfo  {journal} {Science}\ }\textbf {\bibinfo {volume} {358}},\ \bibinfo
  {pages} {79} (\bibinfo {year} {2017})},\ \Eprint
  {http://arxiv.org/abs/https://www.science.org/doi/pdf/10.1126/science.aah6677}
  {https://www.science.org/doi/pdf/10.1126/science.aah6677} \BibitemShut
  {NoStop}%
\bibitem [{\citenamefont {Labzowsky}\ \emph {et~al.}(2007)\citenamefont
  {Labzowsky}, \citenamefont {Schedrin}, \citenamefont {Solovyev},\ and\
  \citenamefont {Plunien}}]{PhysRevLett.98.203003}%
  \BibitemOpen
  \bibfield  {author} {\bibinfo {author} {\bibfnamefont {L.}~\bibnamefont
  {Labzowsky}}, \bibinfo {author} {\bibfnamefont {G.}~\bibnamefont {Schedrin}},
  \bibinfo {author} {\bibfnamefont {D.}~\bibnamefont {Solovyev}}, \ and\
  \bibinfo {author} {\bibfnamefont {G.}~\bibnamefont {Plunien}},\ }\href
  {\doibase 10.1103/PhysRevLett.98.203003} {\bibfield  {journal} {\bibinfo
  {journal} {Phys. Rev. Lett.}\ }\textbf {\bibinfo {volume} {98}},\ \bibinfo
  {pages} {203003} (\bibinfo {year} {2007})}\BibitemShut {NoStop}%
\bibitem [{\citenamefont {Marsman}\ \emph {et~al.}(2015)\citenamefont
  {Marsman}, \citenamefont {Horbatsch},\ and\ \citenamefont
  {Hessels}}]{PhysRevA.91.062506}%
  \BibitemOpen
  \bibfield  {author} {\bibinfo {author} {\bibfnamefont {A.}~\bibnamefont
  {Marsman}}, \bibinfo {author} {\bibfnamefont {M.}~\bibnamefont {Horbatsch}},
  \ and\ \bibinfo {author} {\bibfnamefont {E.~A.}\ \bibnamefont {Hessels}},\
  }\href {\doibase 10.1103/PhysRevA.91.062506} {\bibfield  {journal} {\bibinfo
  {journal} {Phys. Rev. A}\ }\textbf {\bibinfo {volume} {91}},\ \bibinfo
  {pages} {062506} (\bibinfo {year} {2015})}\BibitemShut {NoStop}%
\bibitem [{\citenamefont {Amaro}\ \emph {et~al.}(2018)\citenamefont {Amaro},
  \citenamefont {Loureiro}, \citenamefont {Safari}, \citenamefont {Fratini},
  \citenamefont {Indelicato}, \citenamefont {St\"ohlker},\ and\ \citenamefont
  {Santos}}]{PhysRevA.97.022510}%
  \BibitemOpen
  \bibfield  {author} {\bibinfo {author} {\bibfnamefont {P.}~\bibnamefont
  {Amaro}}, \bibinfo {author} {\bibfnamefont {U.}~\bibnamefont {Loureiro}},
  \bibinfo {author} {\bibfnamefont {L.}~\bibnamefont {Safari}}, \bibinfo
  {author} {\bibfnamefont {F.}~\bibnamefont {Fratini}}, \bibinfo {author}
  {\bibfnamefont {P.}~\bibnamefont {Indelicato}}, \bibinfo {author}
  {\bibfnamefont {T.}~\bibnamefont {St\"ohlker}}, \ and\ \bibinfo {author}
  {\bibfnamefont {J.~P.}\ \bibnamefont {Santos}},\ }\href {\doibase
  10.1103/PhysRevA.97.022510} {\bibfield  {journal} {\bibinfo  {journal} {Phys.
  Rev. A}\ }\textbf {\bibinfo {volume} {97}},\ \bibinfo {pages} {022510}
  (\bibinfo {year} {2018})}\BibitemShut {NoStop}%
\bibitem [{\citenamefont {Amaro}\ \emph
  {et~al.}(2015{\natexlab{a}})\citenamefont {Amaro}, \citenamefont {Franke},
  \citenamefont {Krauth}, \citenamefont {Diepold}, \citenamefont {Fratini},
  \citenamefont {Safari}, \citenamefont {Machado}, \citenamefont {Antognini},
  \citenamefont {Kottmann}, \citenamefont {Indelicato}, \citenamefont {Pohl},\
  and\ \citenamefont {Santos}}]{PhysRevA.92.022514}%
  \BibitemOpen
  \bibfield  {author} {\bibinfo {author} {\bibfnamefont {P.}~\bibnamefont
  {Amaro}}, \bibinfo {author} {\bibfnamefont {B.}~\bibnamefont {Franke}},
  \bibinfo {author} {\bibfnamefont {J.~J.}\ \bibnamefont {Krauth}}, \bibinfo
  {author} {\bibfnamefont {M.}~\bibnamefont {Diepold}}, \bibinfo {author}
  {\bibfnamefont {F.}~\bibnamefont {Fratini}}, \bibinfo {author} {\bibfnamefont
  {L.}~\bibnamefont {Safari}}, \bibinfo {author} {\bibfnamefont
  {J.}~\bibnamefont {Machado}}, \bibinfo {author} {\bibfnamefont
  {A.}~\bibnamefont {Antognini}}, \bibinfo {author} {\bibfnamefont
  {F.}~\bibnamefont {Kottmann}}, \bibinfo {author} {\bibfnamefont
  {P.}~\bibnamefont {Indelicato}}, \bibinfo {author} {\bibfnamefont
  {R.}~\bibnamefont {Pohl}}, \ and\ \bibinfo {author} {\bibfnamefont {J.~P.}\
  \bibnamefont {Santos}},\ }\href {\doibase 10.1103/PhysRevA.92.022514}
  {\bibfield  {journal} {\bibinfo  {journal} {Phys. Rev. A}\ }\textbf {\bibinfo
  {volume} {92}},\ \bibinfo {pages} {022514} (\bibinfo {year}
  {2015}{\natexlab{a}})}\BibitemShut {NoStop}%
\bibitem [{\citenamefont {Udem}\ \emph {et~al.}(2019)\citenamefont {Udem},
  \citenamefont {Maisenbacher}, \citenamefont {Matveev}, \citenamefont
  {Andreev}, \citenamefont {Grinin}, \citenamefont {Beyer}, \citenamefont
  {Kolachevsky}, \citenamefont {Pohl}, \citenamefont {Yost},\ and\
  \citenamefont {Hänsch}}]{https://doi.org/10.1002/andp.201900044}%
  \BibitemOpen
  \bibfield  {author} {\bibinfo {author} {\bibfnamefont {T.}~\bibnamefont
  {Udem}}, \bibinfo {author} {\bibfnamefont {L.}~\bibnamefont {Maisenbacher}},
  \bibinfo {author} {\bibfnamefont {A.}~\bibnamefont {Matveev}}, \bibinfo
  {author} {\bibfnamefont {V.}~\bibnamefont {Andreev}}, \bibinfo {author}
  {\bibfnamefont {A.}~\bibnamefont {Grinin}}, \bibinfo {author} {\bibfnamefont
  {A.}~\bibnamefont {Beyer}}, \bibinfo {author} {\bibfnamefont
  {N.}~\bibnamefont {Kolachevsky}}, \bibinfo {author} {\bibfnamefont
  {R.}~\bibnamefont {Pohl}}, \bibinfo {author} {\bibfnamefont {D.~C.}\
  \bibnamefont {Yost}}, \ and\ \bibinfo {author} {\bibfnamefont {T.~W.}\
  \bibnamefont {Hänsch}},\ }\href {\doibase
  https://doi.org/10.1002/andp.201900044} {\bibfield  {journal} {\bibinfo
  {journal} {Annalen der Physik}\ }\textbf {\bibinfo {volume} {531}},\ \bibinfo
  {pages} {1900044} (\bibinfo {year} {2019})},\ \Eprint
  {http://arxiv.org/abs/https://onlinelibrary.wiley.com/doi/pdf/10.1002/andp.201900044}
  {https://onlinelibrary.wiley.com/doi/pdf/10.1002/andp.201900044} \BibitemShut
  {NoStop}%
\bibitem [{\citenamefont {Matveev}\ \emph {et~al.}(2019)\citenamefont
  {Matveev}, \citenamefont {Kolachevsky}, \citenamefont {Adhikari},\ and\
  \citenamefont {Jentschura}}]{Matveev_2019}%
  \BibitemOpen
  \bibfield  {author} {\bibinfo {author} {\bibfnamefont {A.}~\bibnamefont
  {Matveev}}, \bibinfo {author} {\bibfnamefont {N.}~\bibnamefont
  {Kolachevsky}}, \bibinfo {author} {\bibfnamefont {C.~M.}\ \bibnamefont
  {Adhikari}}, \ and\ \bibinfo {author} {\bibfnamefont {U.~D.}\ \bibnamefont
  {Jentschura}},\ }\href {\doibase 10.1088/1361-6455/ab08e1-1} {\bibfield
  {journal} {\bibinfo  {journal} {Journal of Physics B: Atomic, Molecular and
  Optical Physics}\ }\textbf {\bibinfo {volume} {52}},\ \bibinfo {pages}
  {075006} (\bibinfo {year} {2019})}\BibitemShut {NoStop}%
\bibitem [{\citenamefont {Labzowsky}\ \emph {et~al.}(2002)\citenamefont
  {Labzowsky}, \citenamefont {Soloviev}, \citenamefont {Plunien},\ and\
  \citenamefont {Soff}}]{PhysRevA.65.054502}%
  \BibitemOpen
  \bibfield  {author} {\bibinfo {author} {\bibfnamefont {L.}~\bibnamefont
  {Labzowsky}}, \bibinfo {author} {\bibfnamefont {D.}~\bibnamefont {Soloviev}},
  \bibinfo {author} {\bibfnamefont {G.}~\bibnamefont {Plunien}}, \ and\
  \bibinfo {author} {\bibfnamefont {G.}~\bibnamefont {Soff}},\ }\href {\doibase
  10.1103/PhysRevA.65.054502} {\bibfield  {journal} {\bibinfo  {journal} {Phys.
  Rev. A}\ }\textbf {\bibinfo {volume} {65}},\ \bibinfo {pages} {054502}
  (\bibinfo {year} {2002})}\BibitemShut {NoStop}%
\bibitem [{\citenamefont {Fano}(1961)}]{Fano}%
  \BibitemOpen
  \bibfield  {author} {\bibinfo {author} {\bibfnamefont {U.}~\bibnamefont
  {Fano}},\ }\href {\doibase 10.1103/PhysRev.124.1866} {\bibfield  {journal}
  {\bibinfo  {journal} {Phys. Rev.}\ }\textbf {\bibinfo {volume} {124}},\
  \bibinfo {pages} {1866} (\bibinfo {year} {1961})}\BibitemShut {NoStop}%
\bibitem [{\citenamefont {Bjorken}\ and\ \citenamefont
  {Drell}(1964)}]{Bjorken}%
  \BibitemOpen
  \bibfield  {author} {\bibinfo {author} {\bibfnamefont {J.~D.}\ \bibnamefont
  {Bjorken}}\ and\ \bibinfo {author} {\bibfnamefont {S.~D.}\ \bibnamefont
  {Drell}},\ }\href {https://cds.cern.ch/record/100769} {\emph {\bibinfo
  {title} {{Relativistic quantum mechanics}}}},\ International series in pure
  and applied physics\ (\bibinfo  {publisher} {McGraw-Hill},\ \bibinfo
  {address} {New York, NY},\ \bibinfo {year} {1964})\BibitemShut {NoStop}%
\bibitem [{\citenamefont {de~Beauvoir}\ \emph {et~al.}(1997)\citenamefont
  {de~Beauvoir}, \citenamefont {Nez}, \citenamefont {Julien}, \citenamefont
  {Cagnac}, \citenamefont {Biraben}, \citenamefont {Touahri}, \citenamefont
  {Hilico}, \citenamefont {Acef}, \citenamefont {Clairon},\ and\ \citenamefont
  {Zondy}}]{deBov0}%
  \BibitemOpen
  \bibfield  {author} {\bibinfo {author} {\bibfnamefont {B.}~\bibnamefont
  {de~Beauvoir}}, \bibinfo {author} {\bibfnamefont {F.}~\bibnamefont {Nez}},
  \bibinfo {author} {\bibfnamefont {L.}~\bibnamefont {Julien}}, \bibinfo
  {author} {\bibfnamefont {B.}~\bibnamefont {Cagnac}}, \bibinfo {author}
  {\bibfnamefont {F.}~\bibnamefont {Biraben}}, \bibinfo {author} {\bibfnamefont
  {D.}~\bibnamefont {Touahri}}, \bibinfo {author} {\bibfnamefont
  {L.}~\bibnamefont {Hilico}}, \bibinfo {author} {\bibfnamefont
  {O.}~\bibnamefont {Acef}}, \bibinfo {author} {\bibfnamefont {A.}~\bibnamefont
  {Clairon}}, \ and\ \bibinfo {author} {\bibfnamefont {J.~J.}\ \bibnamefont
  {Zondy}},\ }\href {\doibase 10.1103/PhysRevLett.78.440} {\bibfield  {journal}
  {\bibinfo  {journal} {Phys. Rev. Lett.}\ }\textbf {\bibinfo {volume} {78}},\
  \bibinfo {pages} {440} (\bibinfo {year} {1997})}\BibitemShut {NoStop}%
\bibitem [{\citenamefont {Schwob}\ \emph {et~al.}(1999)\citenamefont {Schwob},
  \citenamefont {Jozefowski}, \citenamefont {{de Beauvoir}}, \citenamefont
  {Hilico}, \citenamefont {Nez}, \citenamefont {Julien}, \citenamefont
  {Biraben}, \citenamefont {Acef}, \citenamefont {Zondy},\ and\ \citenamefont
  {Clairon}}]{deB-1}%
  \BibitemOpen
  \bibfield  {author} {\bibinfo {author} {\bibfnamefont {C.}~\bibnamefont
  {Schwob}}, \bibinfo {author} {\bibfnamefont {L.}~\bibnamefont {Jozefowski}},
  \bibinfo {author} {\bibfnamefont {B.}~\bibnamefont {{de Beauvoir}}}, \bibinfo
  {author} {\bibfnamefont {L.}~\bibnamefont {Hilico}}, \bibinfo {author}
  {\bibfnamefont {F.}~\bibnamefont {Nez}}, \bibinfo {author} {\bibfnamefont
  {L.}~\bibnamefont {Julien}}, \bibinfo {author} {\bibfnamefont
  {F.}~\bibnamefont {Biraben}}, \bibinfo {author} {\bibfnamefont
  {O.}~\bibnamefont {Acef}}, \bibinfo {author} {\bibfnamefont {J.-J.}\
  \bibnamefont {Zondy}}, \ and\ \bibinfo {author} {\bibfnamefont
  {A.}~\bibnamefont {Clairon}},\ }\href {\doibase 10.1103/PhysRevLett.82.4960}
  {\bibfield  {journal} {\bibinfo  {journal} {Phys. Rev. Lett.}\ }\textbf
  {\bibinfo {volume} {82}},\ \bibinfo {pages} {4960} (\bibinfo {year}
  {1999})}\BibitemShut {NoStop}%
\bibitem [{\citenamefont {Yost}\ \emph {et~al.}(2014)\citenamefont {Yost},
  \citenamefont {Matveev}, \citenamefont {Peters}, \citenamefont {Beyer},
  \citenamefont {H\"ansch},\ and\ \citenamefont {Udem}}]{PhysRevA.90.012512}%
  \BibitemOpen
  \bibfield  {author} {\bibinfo {author} {\bibfnamefont {D.~C.}\ \bibnamefont
  {Yost}}, \bibinfo {author} {\bibfnamefont {A.}~\bibnamefont {Matveev}},
  \bibinfo {author} {\bibfnamefont {E.}~\bibnamefont {Peters}}, \bibinfo
  {author} {\bibfnamefont {A.}~\bibnamefont {Beyer}}, \bibinfo {author}
  {\bibfnamefont {T.~W.}\ \bibnamefont {H\"ansch}}, \ and\ \bibinfo {author}
  {\bibfnamefont {T.}~\bibnamefont {Udem}},\ }\href {\doibase
  10.1103/PhysRevA.90.012512} {\bibfield  {journal} {\bibinfo  {journal} {Phys.
  Rev. A}\ }\textbf {\bibinfo {volume} {90}},\ \bibinfo {pages} {012512}
  (\bibinfo {year} {2014})}\BibitemShut {NoStop}%
\bibitem [{\citenamefont {Fleurbaey}\ \emph {et~al.}(2017)\citenamefont
  {Fleurbaey}, \citenamefont {Biraben}, \citenamefont {Julien}, \citenamefont
  {Karr},\ and\ \citenamefont {Nez}}]{PhysRevA.95.052503}%
  \BibitemOpen
  \bibfield  {author} {\bibinfo {author} {\bibfnamefont {H.}~\bibnamefont
  {Fleurbaey}}, \bibinfo {author} {\bibfnamefont {F.}~\bibnamefont {Biraben}},
  \bibinfo {author} {\bibfnamefont {L.}~\bibnamefont {Julien}}, \bibinfo
  {author} {\bibfnamefont {J.-P.}\ \bibnamefont {Karr}}, \ and\ \bibinfo
  {author} {\bibfnamefont {F.}~\bibnamefont {Nez}},\ }\href {\doibase
  10.1103/PhysRevA.95.052503} {\bibfield  {journal} {\bibinfo  {journal} {Phys.
  Rev. A}\ }\textbf {\bibinfo {volume} {95}},\ \bibinfo {pages} {052503}
  (\bibinfo {year} {2017})}\BibitemShut {NoStop}%
\bibitem [{\citenamefont {Fleurbaey}\ \emph {et~al.}(2018)\citenamefont
  {Fleurbaey}, \citenamefont {Galtier}, \citenamefont {Thomas}, \citenamefont
  {Bonnaud}, \citenamefont {Julien}, \citenamefont {Biraben}, \citenamefont
  {Nez}, \citenamefont {Abgrall},\ and\ \citenamefont {Gu\'ena}}]{Fleurbaey}%
  \BibitemOpen
  \bibfield  {author} {\bibinfo {author} {\bibfnamefont {H.}~\bibnamefont
  {Fleurbaey}}, \bibinfo {author} {\bibfnamefont {S.}~\bibnamefont {Galtier}},
  \bibinfo {author} {\bibfnamefont {S.}~\bibnamefont {Thomas}}, \bibinfo
  {author} {\bibfnamefont {M.}~\bibnamefont {Bonnaud}}, \bibinfo {author}
  {\bibfnamefont {L.}~\bibnamefont {Julien}}, \bibinfo {author} {\bibfnamefont
  {F.}~\bibnamefont {Biraben}}, \bibinfo {author} {\bibfnamefont
  {F.}~\bibnamefont {Nez}}, \bibinfo {author} {\bibfnamefont {M.}~\bibnamefont
  {Abgrall}}, \ and\ \bibinfo {author} {\bibfnamefont {J.}~\bibnamefont
  {Gu\'ena}},\ }\href {\doibase 10.1103/PhysRevLett.120.183001} {\bibfield
  {journal} {\bibinfo  {journal} {Phys. Rev. Lett.}\ }\textbf {\bibinfo
  {volume} {120}},\ \bibinfo {pages} {183001} (\bibinfo {year}
  {2018})}\BibitemShut {NoStop}%
\bibitem [{\citenamefont {Solovyev}\ \emph {et~al.}(2020)\citenamefont
  {Solovyev}, \citenamefont {Anikin}, \citenamefont {Zalialiutdinov},\ and\
  \citenamefont {Labzowsky}}]{Solovyev2020}%
  \BibitemOpen
  \bibfield  {author} {\bibinfo {author} {\bibfnamefont {D.}~\bibnamefont
  {Solovyev}}, \bibinfo {author} {\bibfnamefont {A.}~\bibnamefont {Anikin}},
  \bibinfo {author} {\bibfnamefont {T.}~\bibnamefont {Zalialiutdinov}}, \ and\
  \bibinfo {author} {\bibfnamefont {L.}~\bibnamefont {Labzowsky}},\ }\href@noop
  {} {\bibfield  {journal} {\bibinfo  {journal} {Journal of Physics B: Atomic,
  Molecular and Optical Physics}\ }\textbf {\bibinfo {volume} {53}},\ \bibinfo
  {pages} {125002} (\bibinfo {year} {2020})}\BibitemShut {NoStop}%
\bibitem [{\citenamefont {Amaro}\ \emph
  {et~al.}(2015{\natexlab{b}})\citenamefont {Amaro}, \citenamefont {Fratini},
  \citenamefont {Safari}, \citenamefont {Antognini}, \citenamefont
  {Indelicato}, \citenamefont {Pohl},\ and\ \citenamefont
  {Santos}}]{PhysRevA.92.062506}%
  \BibitemOpen
  \bibfield  {author} {\bibinfo {author} {\bibfnamefont {P.}~\bibnamefont
  {Amaro}}, \bibinfo {author} {\bibfnamefont {F.}~\bibnamefont {Fratini}},
  \bibinfo {author} {\bibfnamefont {L.}~\bibnamefont {Safari}}, \bibinfo
  {author} {\bibfnamefont {A.}~\bibnamefont {Antognini}}, \bibinfo {author}
  {\bibfnamefont {P.}~\bibnamefont {Indelicato}}, \bibinfo {author}
  {\bibfnamefont {R.}~\bibnamefont {Pohl}}, \ and\ \bibinfo {author}
  {\bibfnamefont {J.~P.}\ \bibnamefont {Santos}},\ }\href {\doibase
  10.1103/PhysRevA.92.062506} {\bibfield  {journal} {\bibinfo  {journal} {Phys.
  Rev. A}\ }\textbf {\bibinfo {volume} {92}},\ \bibinfo {pages} {062506}
  (\bibinfo {year} {2015}{\natexlab{b}})}\BibitemShut {NoStop}%
\bibitem [{\citenamefont {Anikin}\ \emph {et~al.}(2021)\citenamefont {Anikin},
  \citenamefont {Zalialiutdinov},\ and\ \citenamefont
  {Solovyev}}]{PhysRevA.103.022833}%
  \BibitemOpen
  \bibfield  {author} {\bibinfo {author} {\bibfnamefont {A.}~\bibnamefont
  {Anikin}}, \bibinfo {author} {\bibfnamefont {T.}~\bibnamefont
  {Zalialiutdinov}}, \ and\ \bibinfo {author} {\bibfnamefont {D.}~\bibnamefont
  {Solovyev}},\ }\href {\doibase 10.1103/PhysRevA.103.022833} {\bibfield
  {journal} {\bibinfo  {journal} {Phys. Rev. A}\ }\textbf {\bibinfo {volume}
  {103}},\ \bibinfo {pages} {022833} (\bibinfo {year} {2021})}\BibitemShut
  {NoStop}%
\bibitem [{\citenamefont {Zalialiutdinov}\ \emph {et~al.}(2021)\citenamefont
  {Zalialiutdinov}, \citenamefont {Anikin},\ and\ \citenamefont
  {Solovyev}}]{Zalialiutdinov2021}%
  \BibitemOpen
  \bibfield  {author} {\bibinfo {author} {\bibfnamefont {T.}~\bibnamefont
  {Zalialiutdinov}}, \bibinfo {author} {\bibfnamefont {A.}~\bibnamefont
  {Anikin}}, \ and\ \bibinfo {author} {\bibfnamefont {D.}~\bibnamefont
  {Solovyev}},\ }\href@noop {} {\bibfield  {journal} {\bibinfo  {journal}
  {Journal of Physics B: Atomic, Molecular and Optical Physics}\ }\textbf
  {\bibinfo {volume} {54}},\ \bibinfo {pages} {165002} (\bibinfo {year}
  {2021})}\BibitemShut {NoStop}%
\bibitem [{\citenamefont {Labzowsky}\ \emph {et~al.}(1994)\citenamefont
  {Labzowsky}, \citenamefont {Karasiev},\ and\ \citenamefont
  {Goidenko}}]{Labzowsky1994}%
  \BibitemOpen
  \bibfield  {author} {\bibinfo {author} {\bibfnamefont {L.}~\bibnamefont
  {Labzowsky}}, \bibinfo {author} {\bibfnamefont {V.}~\bibnamefont {Karasiev}},
  \ and\ \bibinfo {author} {\bibfnamefont {I.}~\bibnamefont {Goidenko}},\
  }\href@noop {} {\bibfield  {journal} {\bibinfo  {journal} {Journal of Physics
  B: Atomic, Molecular and Optical Physics}\ }\textbf {\bibinfo {volume}
  {27}},\ \bibinfo {pages} {L439} (\bibinfo {year} {1994})}\BibitemShut
  {NoStop}%
\bibitem [{\citenamefont {Horbatsch}\ and\ \citenamefont
  {Hessels}(2016)}]{HorbHess}%
  \BibitemOpen
  \bibfield  {author} {\bibinfo {author} {\bibfnamefont {M.}~\bibnamefont
  {Horbatsch}}\ and\ \bibinfo {author} {\bibfnamefont {E.~A.}\ \bibnamefont
  {Hessels}},\ }\href {\doibase 10.1103/PhysRevA.93.022513} {\bibfield
  {journal} {\bibinfo  {journal} {Phys. Rev. A}\ }\textbf {\bibinfo {volume}
  {93}},\ \bibinfo {pages} {022513} (\bibinfo {year} {2016})}\BibitemShut
  {NoStop}%
\bibitem [{\citenamefont {Grinin}(2020)}]{ediss27002}%
  \BibitemOpen
  \bibfield  {author} {\bibinfo {author} {\bibfnamefont {A.}~\bibnamefont
  {Grinin}},\ }\emph {\bibinfo {title} {Two-photon frequency comb spectroscopy
  of atomic hydrogen}},\ \href {\doibase 10.5282/edoc.27002} {Ph.D. thesis},\
  \bibinfo  {school} {LMU M\"unchen: Faculty of Physics} (\bibinfo {year}
  {2020})\BibitemShut {NoStop}%
\bibitem [{\citenamefont {Pohl}\ \emph {et~al.}(2010)\citenamefont {Pohl},
  \citenamefont {Antognini}, \citenamefont {Nez}, \citenamefont {Amaro},
  \citenamefont {Biraben}, \citenamefont {Cardoso}, \citenamefont {Covita},
  \citenamefont {Dax}, \citenamefont {Dhawan}, \citenamefont {Fernandes},
  \citenamefont {Giesen}, \citenamefont {Graf}, \citenamefont {H{\"a}nsch},
  \citenamefont {Indelicato}, \citenamefont {Julien}, \citenamefont {Kao},
  \citenamefont {Knowles}, \citenamefont {Le~Bigot}, \citenamefont {Liu},
  \citenamefont {Lopes}, \citenamefont {Ludhova}, \citenamefont {Monteiro},
  \citenamefont {Mulhauser}, \citenamefont {Nebel}, \citenamefont {Rabinowitz},
  \citenamefont {dos Santos}, \citenamefont {Schaller}, \citenamefont
  {Schuhmann}, \citenamefont {Schwob}, \citenamefont {Taqqu}, \citenamefont
  {Veloso},\ and\ \citenamefont {Kottmann}}]{Pohl}%
  \BibitemOpen
  \bibfield  {author} {\bibinfo {author} {\bibfnamefont {R.}~\bibnamefont
  {Pohl}}, \bibinfo {author} {\bibfnamefont {A.}~\bibnamefont {Antognini}},
  \bibinfo {author} {\bibfnamefont {F.}~\bibnamefont {Nez}}, \bibinfo {author}
  {\bibfnamefont {F.~D.}\ \bibnamefont {Amaro}}, \bibinfo {author}
  {\bibfnamefont {F.}~\bibnamefont {Biraben}}, \bibinfo {author} {\bibfnamefont
  {J.~M.~R.}\ \bibnamefont {Cardoso}}, \bibinfo {author} {\bibfnamefont
  {D.~S.}\ \bibnamefont {Covita}}, \bibinfo {author} {\bibfnamefont
  {A.}~\bibnamefont {Dax}}, \bibinfo {author} {\bibfnamefont {S.}~\bibnamefont
  {Dhawan}}, \bibinfo {author} {\bibfnamefont {L.~M.~P.}\ \bibnamefont
  {Fernandes}}, \bibinfo {author} {\bibfnamefont {A.}~\bibnamefont {Giesen}},
  \bibinfo {author} {\bibfnamefont {T.}~\bibnamefont {Graf}}, \bibinfo {author}
  {\bibfnamefont {T.~W.}\ \bibnamefont {H{\"a}nsch}}, \bibinfo {author}
  {\bibfnamefont {P.}~\bibnamefont {Indelicato}}, \bibinfo {author}
  {\bibfnamefont {L.}~\bibnamefont {Julien}}, \bibinfo {author} {\bibfnamefont
  {C.-Y.}\ \bibnamefont {Kao}}, \bibinfo {author} {\bibfnamefont
  {P.}~\bibnamefont {Knowles}}, \bibinfo {author} {\bibfnamefont {E.-O.}\
  \bibnamefont {Le~Bigot}}, \bibinfo {author} {\bibfnamefont {Y.-W.}\
  \bibnamefont {Liu}}, \bibinfo {author} {\bibfnamefont {J.~A.~M.}\
  \bibnamefont {Lopes}}, \bibinfo {author} {\bibfnamefont {L.}~\bibnamefont
  {Ludhova}}, \bibinfo {author} {\bibfnamefont {C.~M.~B.}\ \bibnamefont
  {Monteiro}}, \bibinfo {author} {\bibfnamefont {F.}~\bibnamefont {Mulhauser}},
  \bibinfo {author} {\bibfnamefont {T.}~\bibnamefont {Nebel}}, \bibinfo
  {author} {\bibfnamefont {P.}~\bibnamefont {Rabinowitz}}, \bibinfo {author}
  {\bibfnamefont {J.~M.~F.}\ \bibnamefont {dos Santos}}, \bibinfo {author}
  {\bibfnamefont {L.~A.}\ \bibnamefont {Schaller}}, \bibinfo {author}
  {\bibfnamefont {K.}~\bibnamefont {Schuhmann}}, \bibinfo {author}
  {\bibfnamefont {C.}~\bibnamefont {Schwob}}, \bibinfo {author} {\bibfnamefont
  {D.}~\bibnamefont {Taqqu}}, \bibinfo {author} {\bibfnamefont {J.~F. C.~A.}\
  \bibnamefont {Veloso}}, \ and\ \bibinfo {author} {\bibfnamefont
  {F.}~\bibnamefont {Kottmann}},\ }\href {\doibase 10.1038/nature09250}
  {\bibfield  {journal} {\bibinfo  {journal} {Nature}\ }\textbf {\bibinfo
  {volume} {466}},\ \bibinfo {pages} {213} (\bibinfo {year}
  {2010})}\BibitemShut {NoStop}%
\bibitem [{\citenamefont {Eikema}\ \emph {et~al.}(2001)\citenamefont {Eikema},
  \citenamefont {Walz},\ and\ \citenamefont {H\"ansch}}]{Eikema2001}%
  \BibitemOpen
  \bibfield  {author} {\bibinfo {author} {\bibfnamefont {K.~S.~E.}\
  \bibnamefont {Eikema}}, \bibinfo {author} {\bibfnamefont {J.}~\bibnamefont
  {Walz}}, \ and\ \bibinfo {author} {\bibfnamefont {T.~W.}\ \bibnamefont
  {H\"ansch}},\ }\href {\doibase 10.1103/PhysRevLett.86.5679} {\bibfield
  {journal} {\bibinfo  {journal} {Phys. Rev. Lett.}\ }\textbf {\bibinfo
  {volume} {86}},\ \bibinfo {pages} {5679} (\bibinfo {year}
  {2001})}\BibitemShut {NoStop}%
\bibitem [{\citenamefont {Labzowsky}\ \emph
  {et~al.}(2009{\natexlab{a}})\citenamefont {Labzowsky}, \citenamefont
  {Solovyev},\ and\ \citenamefont {Plunien}}]{LSP-separation}%
  \BibitemOpen
  \bibfield  {author} {\bibinfo {author} {\bibfnamefont {L.}~\bibnamefont
  {Labzowsky}}, \bibinfo {author} {\bibfnamefont {D.}~\bibnamefont {Solovyev}},
  \ and\ \bibinfo {author} {\bibfnamefont {G.}~\bibnamefont {Plunien}},\ }\href
  {\doibase 10.1103/PhysRevA.80.062514} {\bibfield  {journal} {\bibinfo
  {journal} {Phys. Rev. A}\ }\textbf {\bibinfo {volume} {80}},\ \bibinfo
  {pages} {062514} (\bibinfo {year} {2009}{\natexlab{a}})}\BibitemShut
  {NoStop}%
\bibitem [{\citenamefont {Labzowsky}\ \emph
  {et~al.}(2009{\natexlab{b}})\citenamefont {Labzowsky}, \citenamefont
  {Schedrin}, \citenamefont {Solovyev}, \citenamefont {Chernovskaya},
  \citenamefont {Plunien},\ and\ \citenamefont
  {Karshenboim}}]{PhysRevA.79.052506}%
  \BibitemOpen
  \bibfield  {author} {\bibinfo {author} {\bibfnamefont {L.}~\bibnamefont
  {Labzowsky}}, \bibinfo {author} {\bibfnamefont {G.}~\bibnamefont {Schedrin}},
  \bibinfo {author} {\bibfnamefont {D.}~\bibnamefont {Solovyev}}, \bibinfo
  {author} {\bibfnamefont {E.}~\bibnamefont {Chernovskaya}}, \bibinfo {author}
  {\bibfnamefont {G.}~\bibnamefont {Plunien}}, \ and\ \bibinfo {author}
  {\bibfnamefont {S.}~\bibnamefont {Karshenboim}},\ }\href {\doibase
  10.1103/PhysRevA.79.052506} {\bibfield  {journal} {\bibinfo  {journal} {Phys.
  Rev. A}\ }\textbf {\bibinfo {volume} {79}},\ \bibinfo {pages} {052506}
  (\bibinfo {year} {2009}{\natexlab{b}})}\BibitemShut {NoStop}%
\bibitem [{\citenamefont {Bydder}\ \emph {et~al.}(2007)\citenamefont {Bydder},
  \citenamefont {Rahal}, \citenamefont {Fullerton},\ and\ \citenamefont
  {Bydder}}]{Bydder}%
  \BibitemOpen
  \bibfield  {author} {\bibinfo {author} {\bibfnamefont {M.}~\bibnamefont
  {Bydder}}, \bibinfo {author} {\bibfnamefont {A.}~\bibnamefont {Rahal}},
  \bibinfo {author} {\bibfnamefont {G.~D.}\ \bibnamefont {Fullerton}}, \ and\
  \bibinfo {author} {\bibfnamefont {G.~M.}\ \bibnamefont {Bydder}},\ }\href
  {\doibase https://doi.org/10.1002/jmri.20850} {\bibfield  {journal} {\bibinfo
   {journal} {Journal of Magnetic Resonance Imaging}\ }\textbf {\bibinfo
  {volume} {25}},\ \bibinfo {pages} {290} (\bibinfo {year} {2007})},\ \Eprint
  {http://arxiv.org/abs/https://onlinelibrary.wiley.com/doi/pdf/10.1002/jmri.20850}
  {https://onlinelibrary.wiley.com/doi/pdf/10.1002/jmri.20850} \BibitemShut
  {NoStop}%
\bibitem [{\citenamefont {Horbatsch}\ and\ \citenamefont
  {Hessels}(2010)}]{PhysRevA.82.052519}%
  \BibitemOpen
  \bibfield  {author} {\bibinfo {author} {\bibfnamefont {M.}~\bibnamefont
  {Horbatsch}}\ and\ \bibinfo {author} {\bibfnamefont {E.~A.}\ \bibnamefont
  {Hessels}},\ }\href {\doibase 10.1103/PhysRevA.82.052519} {\bibfield
  {journal} {\bibinfo  {journal} {Phys. Rev. A}\ }\textbf {\bibinfo {volume}
  {82}},\ \bibinfo {pages} {052519} (\bibinfo {year} {2010})}\BibitemShut
  {NoStop}%
\bibitem [{\citenamefont {Horbatsch}\ and\ \citenamefont
  {Hessels}(2011)}]{PhysRevA.84.032508}%
  \BibitemOpen
  \bibfield  {author} {\bibinfo {author} {\bibfnamefont {M.}~\bibnamefont
  {Horbatsch}}\ and\ \bibinfo {author} {\bibfnamefont {E.~A.}\ \bibnamefont
  {Hessels}},\ }\href {\doibase 10.1103/PhysRevA.84.032508} {\bibfield
  {journal} {\bibinfo  {journal} {Phys. Rev. A}\ }\textbf {\bibinfo {volume}
  {84}},\ \bibinfo {pages} {032508} (\bibinfo {year} {2011})}\BibitemShut
  {NoStop}%
\bibitem [{\citenamefont {Sansonetti}\ \emph {et~al.}(2011)\citenamefont
  {Sansonetti}, \citenamefont {Simien}, \citenamefont {Gillaspy}, \citenamefont
  {Tan}, \citenamefont {Brewer}, \citenamefont {Brown}, \citenamefont {Wu},\
  and\ \citenamefont {Porto}}]{PhysRevLett.107.023001}%
  \BibitemOpen
  \bibfield  {author} {\bibinfo {author} {\bibfnamefont {C.~J.}\ \bibnamefont
  {Sansonetti}}, \bibinfo {author} {\bibfnamefont {C.~E.}\ \bibnamefont
  {Simien}}, \bibinfo {author} {\bibfnamefont {J.~D.}\ \bibnamefont
  {Gillaspy}}, \bibinfo {author} {\bibfnamefont {J.~N.}\ \bibnamefont {Tan}},
  \bibinfo {author} {\bibfnamefont {S.~M.}\ \bibnamefont {Brewer}}, \bibinfo
  {author} {\bibfnamefont {R.~C.}\ \bibnamefont {Brown}}, \bibinfo {author}
  {\bibfnamefont {S.}~\bibnamefont {Wu}}, \ and\ \bibinfo {author}
  {\bibfnamefont {J.~V.}\ \bibnamefont {Porto}},\ }\href {\doibase
  10.1103/PhysRevLett.107.023001} {\bibfield  {journal} {\bibinfo  {journal}
  {Phys. Rev. Lett.}\ }\textbf {\bibinfo {volume} {107}},\ \bibinfo {pages}
  {023001} (\bibinfo {year} {2011})}\BibitemShut {NoStop}%
\bibitem [{\citenamefont {Labzowsky}\ \emph {et~al.}(1993)\citenamefont
  {Labzowsky}, \citenamefont {Klimchitskaya},\ and\ \citenamefont
  {Dmitriev}}]{LabKlim}%
  \BibitemOpen
  \bibfield  {author} {\bibinfo {author} {\bibfnamefont {L.}~\bibnamefont
  {Labzowsky}}, \bibinfo {author} {\bibfnamefont {G.}~\bibnamefont
  {Klimchitskaya}}, \ and\ \bibinfo {author} {\bibfnamefont {Y.}~\bibnamefont
  {Dmitriev}},\ }\href@noop {} {\emph {\bibinfo {title} {Relativistic Effects
  in the Spectra of Atomic Systems}}}\ (\bibinfo  {publisher} {Institute of
  Physics Publishing},\ \bibinfo {year} {1993})\BibitemShut {NoStop}%
\bibitem [{\citenamefont {Zalialiutdinov}\ \emph {et~al.}(2014)\citenamefont
  {Zalialiutdinov}, \citenamefont {Baukina}, \citenamefont {Solovyev},\ and\
  \citenamefont {Labzowsky}}]{Zalialiutdinov_2014}%
  \BibitemOpen
  \bibfield  {author} {\bibinfo {author} {\bibfnamefont {T.}~\bibnamefont
  {Zalialiutdinov}}, \bibinfo {author} {\bibfnamefont {Y.}~\bibnamefont
  {Baukina}}, \bibinfo {author} {\bibfnamefont {D.}~\bibnamefont {Solovyev}}, \
  and\ \bibinfo {author} {\bibfnamefont {L.}~\bibnamefont {Labzowsky}},\ }\href
  {\doibase 10.1088/0953-4075/47/11/115007} {\bibfield  {journal} {\bibinfo
  {journal} {Journal of Physics B: Atomic, Molecular and Optical Physics}\
  }\textbf {\bibinfo {volume} {47}},\ \bibinfo {pages} {115007} (\bibinfo
  {year} {2014})}\BibitemShut {NoStop}%
\bibitem [{\citenamefont {Varshalovich}\ \emph {et~al.}(1988)\citenamefont
  {Varshalovich}, \citenamefont {Moskalev},\ and\ \citenamefont
  {Khersonskii}}]{VMK}%
  \BibitemOpen
  \bibfield  {author} {\bibinfo {author} {\bibfnamefont {D.~A.}\ \bibnamefont
  {Varshalovich}}, \bibinfo {author} {\bibfnamefont {A.~N.}\ \bibnamefont
  {Moskalev}}, \ and\ \bibinfo {author} {\bibfnamefont {V.~K.}\ \bibnamefont
  {Khersonskii}},\ }\href {\doibase 10.1142/0270} {\emph {\bibinfo {title}
  {Quantum Theory of Angular Momentum}}}\ (\bibinfo  {publisher} {WORLD
  SCIENTIFIC},\ \bibinfo {year} {1988})\ \Eprint
  {http://arxiv.org/abs/https://www.worldscientific.com/doi/pdf/10.1142/0270}
  {https://www.worldscientific.com/doi/pdf/10.1142/0270} \BibitemShut {NoStop}%
\end{thebibliography}%

\appendix
\renewcommand{\theequation}{A\arabic{equation}}
\setcounter{equation}{0}
\setcounter{figure}{0}
\renewcommand{\thefigure}{A\arabic{figure}}
\renewcommand{\thetable}{A\arabic{table}}

\onecolumngrid

\section{Determination of the transition frequency through two-photon scattering}
\subsection{Quantum interference effect}
\label{SuppM2}
The details of the analysis below can be found in \cite{Solovyev2020}. To be self-consistent, we give only a brief summary of it. To determine the resonant transition frequency we use the definition Eq.~(\ref{S2.10}). In case of two-photon scattering process, the dependence of the cross section $\sum\limits_{\vec{e}_f}\sigma_{if}$ on the incident photon frequency represents the natural line profile for the transition $n_i l_i j_i F_i \rightarrow n_r l_r j_r F_r$, which can be expressed by (Fano profile)
\begin{eqnarray}
\label{main}
\sigma_{if}=C\left[\frac{f_{\rm res}}{(\omega_0-\omega)^2+\frac{\Gamma^2}{4}}
+2\mathrm{Re}\frac{f_{\rm nr}}{(\omega_0-\omega-\frac{\mathrm{i}\Gamma}{2})\Delta} \right]
d\omega,
\end{eqnarray}
where $\Delta=E_{n_rl_rj_{r'}F_{r'}}-E_{n_rl_rj_rF_r} $, $\sigma_{if}=\sum\limits_{\vec{e}_f}\sigma_{if}$, $ C $ is some constant that is not important for our further derivations and
\begin{eqnarray}
\label{final}
f_{\rm res}=
\sum_{xy}A_{xy}^{\mathrm{res}}\left\lbrace\left\lbrace e_{1}^{i}\otimes \nu_{1}^{f} \right\rbrace_{y} 
\otimes
\left\lbrace e_{1}^{i}\otimes \nu_{1}^{f} \right\rbrace_{y}\right\rbrace_{00},
\end{eqnarray}
\begin{eqnarray}
\label{final2}
f_{\rm nr}=
\sum_{xy}A_{xy}^{\mathrm{nr}}\left\lbrace\left\lbrace e_{1}^{i}\otimes \nu_{1}^{f} \right\rbrace_{y} 
\otimes
\left\lbrace e_{1}^{i}\otimes \nu_{1}^{f} \right\rbrace_{y}\right\rbrace_{00},
\end{eqnarray}
where $e^i_1$ denotes the polarization vector of the incident photon and $\nu^f_1$ is the direction vector of the outgoing photon.
Coefficients $A_{xy}$ are defined as
\begin{eqnarray}
\label{S31}
A_{xy}^{\mathrm{res}}=\frac{6(-1)^{-y}}{2F_{i}+1}\Pi_{x}^2\Pi_{y}
\begin{Bmatrix}
1 & 1 & y \\
1 & 1 & x
\end{Bmatrix}
\begin{Bmatrix}
1 & 1 & x \\
1 & 1 & 1
\end{Bmatrix}
\begin{Bmatrix}
1 & x & 1 \\
F_r & F_{i} & F_r
\end{Bmatrix}
\begin{Bmatrix}
1 & x & 1 \\
F_r & F_{f} & F_r
\end{Bmatrix}\times
\\
\nonumber
\left|\langle n_i l_i j_i F_i ||r||n_r l_r j_r F_r \rangle 
\langle n_r l_r j_r F_r ||r||n_f l_f j_f F_f \rangle \right|^2,
\end{eqnarray}
\begin{eqnarray}
\label{S32}
A_{xy}^{\mathrm{nr}}=\frac{6(-1)^{F'-F-y}}{2F_{i}+1}\Pi_{x}^2\Pi_{y}
\begin{Bmatrix}
1 & 1 & y \\
1 & 1 & x
\end{Bmatrix}
\begin{Bmatrix}
1 & 1 & x \\
1 & 1 & 1
\end{Bmatrix}
\begin{Bmatrix}
1 & x & 1 \\
F' & F_{i} & F
\end{Bmatrix}
\begin{Bmatrix}
1 & x & 1 \\
F' & F_{f} & F
\end{Bmatrix}\times\qquad
\\
\nonumber
\langle n_i l_i j_i F_i ||r||n_r l_r j_r F_r \rangle 
\langle n_r l_r j_{r'} F_{r'} ||r||n_i l_i j_i F_i\rangle 
\langle n_f l_f j_f F_f||r||n_r l_r j_{r'} F_{r'}\rangle 
\langle n_r l_r j_r F_r ||r||n_f l_f j_f F_f \rangle.
\end{eqnarray}

Using the definition of transition frequency via the "maximum" of the line profile (most probable), Eq.~(\ref{main}), according to Eq.~(\ref{S2.10}), we find 
\begin{eqnarray}
\label{add1}
\frac{d}{d\omega}\sigma_{if}(\omega)
=
-\frac{8\left[f_{\mathrm{nr}}\left(\Gamma^2-4 (\omega-\omega_{0})^2
\right)+4 \Delta  f_{\mathrm{res}} (\omega-\omega_{0})
\right]
}
{\Delta  \left(\Gamma^2+4 (\omega-\omega_{0})^2\right)^2}
=0
.
\end{eqnarray}
Expansion of Eq.~(\ref{add1}) into the Taylor series in the vicinity of $\omega_0$
yields
\begin{eqnarray}
\label{add2}
-\frac{8 f_{\mathrm{nr}}}{\Gamma^2\Delta}
-\frac{32 f_{\mathrm{res}}(\omega -\omega_{0})}{\Gamma^4}
+O\left((\omega -\omega_{0})^2\right)
=0
.
\end{eqnarray}
Finally, neglecting the terms of the order $O\left((\omega -\omega_{0})^2\right)$ in Eq.~(\ref{add2}) and solving it with respect to $\omega$ we arrive at the definition of $\omega_{\rm max}$:
\begin{eqnarray}
\label{23}
\omega_{\rm max}=\omega_{0}-\delta\omega,
\end{eqnarray}
where
\begin{eqnarray}
\label{24}
\delta\omega = \frac{f_{\rm nr}}{f_{\rm res}}\frac{\Gamma^2}{4\Delta}.
\end{eqnarray}
With our definition of $\Delta$ this value corresponds to the lower component of the fine structure of the level $n_rl_r$. For the upper sublevel of two neighboring components of energy level we would arrive to the same expression as Eq.~(\ref{23}) but with the opposite sign of $\Delta$, additional weighting factor originating from the summation over projections in the resonant term, and with $\Gamma=\Gamma_{nlj'F'}$. NR correction in Eq.~(\ref{23}) can depend on the arrangement of the experiment, i.e., on the angle between the vectors $\vec{e}_i$ and $\vec{\nu}_f$. 

The smallness of NR corrections in Eq.~(\ref{24}) is defined by the ratio $\Gamma /\Delta$. Eq.~(\ref{24}) is obtained as the lowest term of the series expansion over $ \Gamma /\Delta $. The approximations that were used for derivation of Eq.~(\ref{24}) are valid up to the higher order terms in parameter $\Gamma /\Delta$. This parameter is always small for two neighbouring components of the fine structure. The parameter $\Gamma /\Delta$ may not be small for two neighbouring hyperfine sublevels, but this requires special study \cite{PhysRevA.79.052506}.

\subsection{Application to $ 2s_{1/2}^{F=0}\rightarrow 4p_{1/2}^{F=1} $ and $ 2s_{1/2}^{F=0}\rightarrow 4p_{3/2}^{F=1} $ transitions}

Now we turn to evaluation of $ 2s_{1/2}^{F=0}\rightarrow 4p_{1/2}^{F=1} $ transition frequency with account for NR corrections originating from the neighboring $ 4p_{3/2}^{F=1} $ level. For this purpose we set in all equations $ n_i l_i=2s $, $ j_i=1/2 $, $ F_i=0 $, $ n_rl_r=4p $, $ j_r=1/2 $, $ F_r=1 $, $ j_{r'}=3/2 $, $ F_{r'}=1 $. For the final states, we chose the states listed in Table~\ref{tab1}. Note that hyperfine structure of $ 1s $ and $ 2s $ electron shells was resolvable in experiments \cite{Science}. The results of evaluations are presented in Table~\ref{tab1}. 
\begin{table}[h]
\renewcommand{\arraystretch}{1.2}
\caption{The NR corrections in kHz to the transitions frequency $ 2s_{1/2}^{F=0}\rightarrow 4p_{1/2}^{F=1} $ with the account for the neighbouring $ 4p_{3/2}^{F=1} $ for the experiment with correlation $\{\vec{e}_i,\vec{\nu}_f\}$.}
\begin{tabular}{|c | c| c |}
\hline
Final state & $\delta\omega$ to $\nu_{1/2}$ \cite{Science} & $\delta\omega$ to $\nu_{3/2}$ \cite{Science}\\
\hline
$1s_{1/2}^{F=0}$ &   60.7127 & -15.1782 \\
\hline
$1s_{1/2}^{F=1}$ &  -30.3564 & 30.3564 \\
\hline
$2s_{1/2}^{F=0}$ &   60.7127 & -15.1782 \\
\hline
$2s_{1/2}^{F=1}$ &  -30.3564 & 30.3564 \\
\hline
$3s_{1/2}^{F=0}$ &   60.7127 & -15.1782 \\
\hline
$3s_{1/2}^{F=1}$ &  -30.3564 & 30.3564 \\
\hline
$3d_{3/2}^{F=1}$ &  -30.3564 & 30.3564 \\
\hline
$3d_{3/2}^{F=2}$ &    6.0713 & -151.7819 \\
\hline
\end{tabular}
\label{tab1}
\end{table}

For evaluation of NR corrections according to Eqs.~(\ref{main}), (\ref{final}), (\ref{24}) we use theoretical values given in \cite{HorbHess}, which incorporate relativistic, QED, nuclear size, the hyperfine structure corrections. Thus, $\Delta = E_{4p_{3/2}^{F=1}}-E_{4p_{1/2}^{F=1}}= 1367433.3$ kHz and the calculated value of the level width is found as $ \Gamma=\Gamma_{4p_{1/2}^{F=1}}=1.2941\times 10^{7} $ Hz. These values give a sufficiently accurate result for $ \delta \omega $ up to four digits after the decimal point. The parameter $ \Gamma /\Delta $ in this case is equal to $ 0.00946 $, so the expansion in powers of this parameter works very well.

As was found in \cite{Solovyev2020}, NR corrections to the transition frequency $2s_{1/2}^{F=0}\rightarrow 4p_{1/2}^{F=1}$ are independent of the experiment "geometry" when the final state is fixed. However, these NR corrections appear to be strongly dependent on the method of frequency registration, i.e., on the choice of the state into which the final excited level decays $4p_{1/2}^{F=1}$. Moreover, this dependence concerns only the quantum numbers of the final state, and the result does not depend on the frequency of the outgoing photon. The latter circumstance is understandable, since according to Eq.~(\ref{24}) the NR corrections are proportional to the ratio $ f_{\rm nr}/f_{\rm res}$, where the corresponding energy differences are reduced. 

When the hyperfine structure of the final levels is resolved, the NR corrections differ only by the values of the total momentum $F_f$ of the final hyperfine sublevel. This can be seen from the closed expressions (\ref{S31}), (\ref{S32}) for the NR corrections via $ 6j $-symbols. Therefore, for the transition frequency $ 2s_{1/2}^{F=0}\rightarrow 4p_{1/2}^{F=1} $, three different values of $ \omega_{\rm res}^{\rm max} $ corresponding to $ F_f=0,\,1,\,2 $ can be derived by using $ \omega_0 $ from \cite{HorbHess} and NR corrections from Table~\ref{tab1}:
\begin{eqnarray}
\label{three}
F_f=0\;\;\;\omega_{\rm res}^{\rm max}
=616520152619.2\;\mathrm{kHz}
\\\nonumber
F_f=1\;\;\;\omega_{\rm res}^{\rm max}
=616520152528.1\;\mathrm{kHz}
\\\nonumber
F_f=2\;\;\;\omega_{\rm res}^{\rm max}
=616520152564.2\;\mathrm{kHz}
\end{eqnarray}

If in the process of the frequency measurement only the emission of the outgoing photon is detected without fixing of its frequency, the summation over all the final states should be done. In the case of our interests this summation looks as follows
\begin{eqnarray}
\label{avr}
\delta\omega = \frac{\sum\limits_{n_fl_fj_fF_f}f_{\rm nr}}{\sum\limits_{n_fl_fj_fF_f}f_{\rm res}}\frac{\Gamma^2}{4\Delta}
.
\end{eqnarray}
Now the NR correction begins to depend on the angle between the vectors $ \vec{e}_i $, $ \vec{\nu}_f $, see \cite{Science,Solovyev2020}.

According to Eq.~(\ref{avr}), the NR correction vanishes for certain angles $\theta_1=\arccos(1/\sqrt{3}) = 54.7^{\circ}$ and $\theta_2=\pi-\theta_1 = 125.3^{\circ}$. The possibility of using "magic angles" to determine the transition frequencies in atoms was mentioned in \cite{HorbHess,https://doi.org/10.1002/andp.201900044}. In \cite{https://doi.org/10.1002/andp.201900044} it was noted that the method of extracting the transition frequency value from the experimental data used in \cite{Science} is actually equivalent to the use of "magic angles". The same "magic angles" arise in different areas of quantum physics where the interference of two electric dipole amplitudes is involved, see for example \cite{Bydder}.

Recently, the evaluation of atomic transition frequencies using "magic angles" was considered in \cite{PhysRevA.92.062506}. The values of "magic angles" in \cite{PhysRevA.92.062506} coincide with those given above for similar transitions. Evaluating the transition frequencies $2s_{1/2}^{F=0}-4p_{1/2}^{F=1}$ and $2s_{1/2}^{F=0}-4p_{3/2}^{F=1}$ using the equation (\ref{avr}) for "magic angles" with theoretical values $\omega_0$, $\Delta$ and $\Gamma$ gives
\begin{eqnarray}
\label{new1}
\omega_{\rm res}^{\rm max}=616520152558.5\;\mathrm{kHz},
\\
\nonumber
\omega_{\rm res}^{\rm max}=616521519991.8\;\mathrm{kHz},
\end{eqnarray}
which are within the experimental error bars reported in \cite{Science}. Additional mathematical details to derive the above formulas can be found in \cite{Solovyev2020}.

\renewcommand{\theequation}{B\arabic{equation}}
\setcounter{equation}{0}
\setcounter{figure}{0}
\renewcommand{\thefigure}{B\arabic{figure}}
\renewcommand{\thetable}{B\arabic{table}}

\section{Quantum interference effect in a cascade process}

\subsection{Four-photon scattering process with one-photon absorption and three-photon cascade emission}
\label{SuppM1}
The theoretical derivation of the line profile is usually given by considering the photon scattering process within the resonance approximation \cite{ANDREEV2008135}. Within this approach, only the part that corresponds to the resonant process is studied, while the rest is discarded. Then, since the absorption part and the part responsible for the emission process are given as the numerator of the scattering amplitude, the resulting profile can be attributed with equal right to both absorption and emission. The dominant nonresonant effect arises due to the state adjacent to the resonant one \cite{Jent-NR}. The idea of \cite{Jent-NR} is widely utilized in modern studies of the photon scattering processes on atoms, see, for example, \cite{PhysRevA.82.052519,PhysRevA.84.032508,PhysRevLett.107.023001,PhysRevA.92.022514,PhysRevA.92.062506,PhysRevA.97.022510}. The various features of nonresonant contributions are of particular importance for precision spectroscopic experiments such as \cite{Science}. For example, the attention of researchers was drawn to the angular dependence of the resulting corrections to the transition frequency. As a rule, the subject of study is the process of two-photon scattering, when one photon is absorbed and the other is emitted. Theoretical calculations of NR corrections for the experiments based on two-photon spectroscopy (two photons are absorbed) were presented recently in \cite{PhysRevA.103.022833,Zalialiutdinov2021}, where the dependence on the external experimental conditions was also discussed on example of helium atom \cite{Zalialiutdinov2021}.  The main conclusion of all such theoretical calculations require special consideration of nonresonant corrections for each particular experiment. 

\begin{figure}[hbtp]
	\centering
	\includegraphics[scale=0.25]{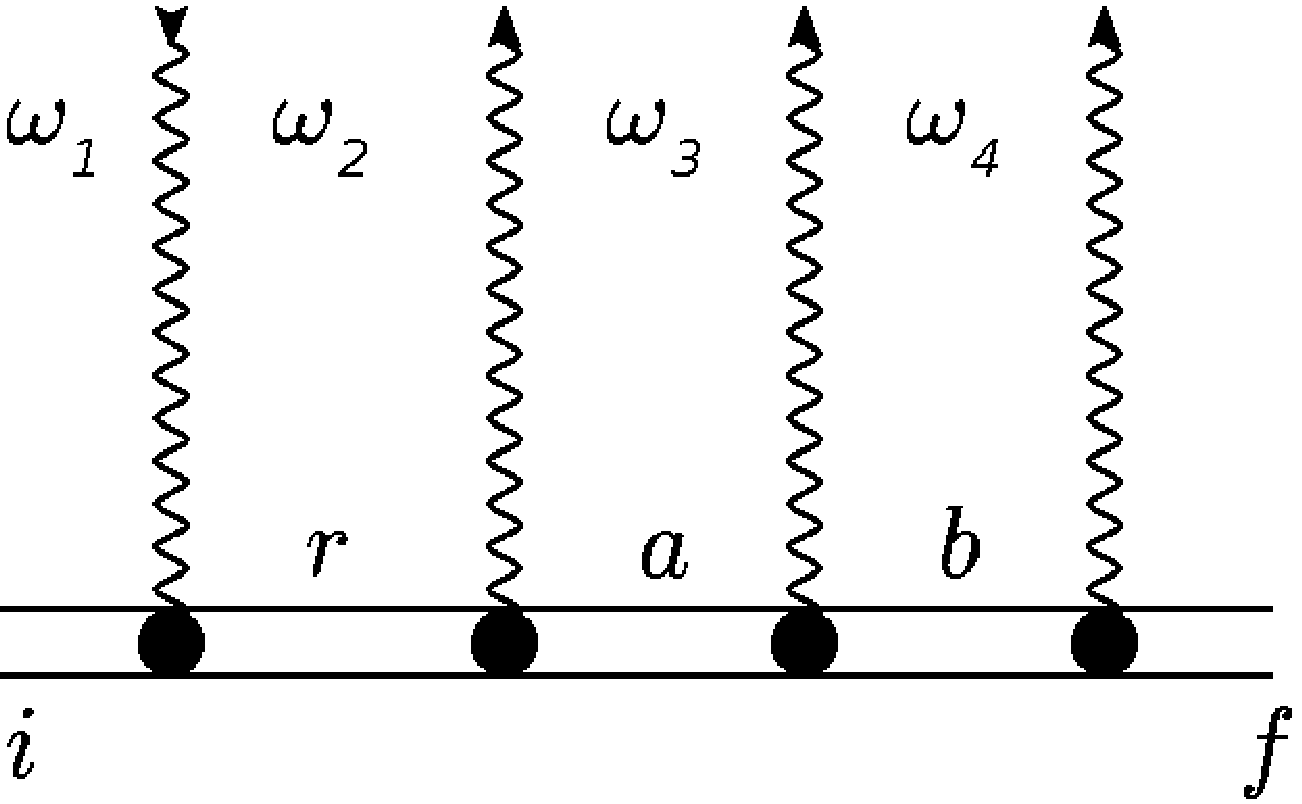}
	\caption{A Feynman graph depicting a four-photon scattering process with one-photon absorption link. Here $i$, $f$ denote the initial and final states, respectively, $r$, $a$ and $b$ are the intermediate resonant states. State $r$ is the resonant contribution to the absorbing photons, $a$ and $b$ reflect the cascade contribution in the radiation process. The frequencies of the emitted photon is denoted by $\omega_2$, $\omega_3$, $\omega_4$ and the absorbed photon $\omega_1$.}\label{Fig_S2}
\end{figure}

Becoming an inevitable part of precision atomic physics, nonresonant effects arising in the transition frequency measurement process can play a decisive role. Such effects, and in particular QIE (quantum interference effect), should obligatory be taken into account in experiments pursuing the goal of increasing the accuracy. In this connection the detailed theoretical description of the experiment should consider the accompanying details of the process beyond the resonance approximation. For example, it can be shown that in four-photon scattering process (one photon is absorbed and three are emitted) in experiments like \cite{Science} the cascade radiation affects the absorption profile. As mentioned above, this possible effect follows directly from the theoretical derivation of the line profile. To demonstrate this explicitly, here we consider the influence of interfering paths in cascade radiation. The interference effect arises with description of the process $i+\gamma\rightarrow r \rightarrow a +\gamma  \rightarrow b+\gamma\rightarrow f+\gamma$, where$\gamma$ is the absorbed or emitted photon, $i$ and $f$ denote initial and final atomic states, respectively, $r$ corresponds to the resonant state under the study, and $a$, $b$ represent the states corresponding to the cascade in radiation. According to theoretical foundations, the very construction of the QED theory requires a description of the photon scattering by atoms, starting from the (meta)stable state and ending with the (meta)stable atomic level. Thus, in further we restrict our selves by the consideration of the initial state $i=2s_{1/2}^{F=0}$ and the final state $1s_{1/2}$. As a first step, we use the resonance approximation only to describe the cascade radiation, assuming the smallness of the effects beyond it, see, for example, the problem discussed in \cite{LSP-separation}. Schematically, such a description can be illustrated by the Feynman graph in Fig.~\ref{Fig_S2}, where the initial and final states are assumed to be (meta)stable, and the process occurs through one-photon absorption to the resonance state $r$ culminating in cascade radiation with the resonant states $a$ and $b$.

The resonant approximation adopted for cascade radiation means that the Feynman graphs accounting for photon permutations are omitted in Fig.~\ref{Fig_S2} and only the resonance terms are left in the arising sums over the entire spectrum (see below). Then, the $S$-matrix element (in relativistic units $\hbar=c=1$) corresponding to the diagram in Fig.~\ref{Fig_S2} is 
\begin{eqnarray}
\label{S2.1}
S^{(4\gamma)}_{fi} = (-i e)^4\int d^4x_1\,d^4x_2\,d^4x_3\,d^4x_4\overline{\psi}_f(x_1)\left(\gamma^{\mu_1} A_{\mu_1}^*(x_1)\right)S(x_1,x_2)\left(\gamma^{\mu_2} A_{\mu_2}^*(x_2)\right)\times
\\
\nonumber
S(x_2,x_3)\left(\gamma^{\mu_3} A^*_{\mu_3}(x_3)\right)S(x_3,x_4)\left(\gamma^{\mu_4} A_{\mu_4}(x_4)\right)\psi_i(x_4).
\end{eqnarray}

For an arbitrary atomic state $A$
\begin{eqnarray}
\label{S2.2}
\psi_A(x)=\psi_A(\vec{r})e^{-iE_At},
\end{eqnarray}
where $\psi_A(\vec{r})$ is the solution of the Dirac equation for the atomic electron, $E_A$ is the Dirac energy, $\overline{\psi}_A=\psi^+_A\gamma_0$ is the Dirac conjugated wave function, $\gamma_{\mu}\equiv(\gamma_0,\vec{\gamma})$ are the Dirac matrices and $x\equiv(t,\vec{r})$ is the four-dimensional space-time coordinate. 
The photon field or the photon wave function $A_{\mu}(x)$ is defined by
\begin{eqnarray}
\label{S2.3}
A_{\mu}(x)=\sqrt{\frac{2\pi}{\omega}}e_{\mu}e^{i(\vec{k}\vec{r}-\omega t)}=e^{-i\omega t}A_{\mu}(\vec{r}),
\end{eqnarray}
where $e_{\mu}$ are the components of the photon polarization four-vector ($\vec{e}$ is 3-dimensional polarization vector for real photons), $k\equiv(\omega,\vec{k})$ is the photon momentum four-vector, $\vec{k}$ is the wave vector, $\omega=|\vec{k}|$ is the photon frequency. Eq.~(\ref{S2.3}) corresponds to the absorbed photon and $A^*_{\mu}(x)$ corresponds to the emitted photon. Finally, the electron propagator for the bound electron can be presented in the form of the eigenmode decomposition with respect to one-electron eigenstates:
\begin{eqnarray}
\label{S2.4}
S(x_1,x_2)=\frac{i}{2\pi}\int\limits^{\infty}_{-\infty}d\Omega\, e^{-i\Omega(t_1-t_2)}\sum\limits_{n}\frac{\psi_n(\vec{r}_1)\overline{\psi}_n(\vec{r}_2)}{\Omega-E_n(1-i0)},
\end{eqnarray}
where the sum over $n$ runs the entire Dirac spectrum.

By integrating over time variables, four $\delta$-functions can be obtained: $(2\pi)^4\delta(E_f+\omega_4-\Omega_1)\delta(\Omega_1+\omega_3-\Omega_2)\delta(\Omega_2+\omega_2-\Omega_3)\delta(\Omega_3-\omega_1-E_i)$. Then integration over $\Omega_i$ reduces them to a $\delta$-function representing the conservation law of the four-photon scattering process, leading to
\begin{eqnarray}
\label{S2.5}
S^{(4\gamma)}_{fi} = - 2\pi i e^4 \delta\left(E_f-\omega_1+\omega_2+\omega_3+\omega_4-E_i\right)\int d^3r_1\,d^3r_2\,d^3r_3\,d^3r_4\, \overline{\psi}_f(\vec{r}_1)\times\qquad
\\
\nonumber
\left(\gamma^{\mu_1} A_{\mu_1}^*(\vec{r}_1)\right)\sum\limits_{n_1}\frac{\psi_{n_1}(\vec{r}_1)\overline{\psi}_{n_1}(\vec{r}_2)}{E_f+\omega_4-E_{n_1}(1-i0)}
\left(\gamma^{\mu_2} A_{\mu_2}^*(\vec{r}_2)\right)
\sum\limits_{n_2}\frac{\psi_{n_2}(\vec{r}_2)\overline{\psi}_{n_2}(\vec{r}_3)}{E_f+\omega_3+\omega_4-E_{n_2}(1-i0)}
\\
\nonumber
\times\left(\gamma^{\mu_3} A_{\mu_3}^*(\vec{r}_3)\right)\sum\limits_{n_3}\frac{\psi_{n_3}(\vec{r}_3)\overline{\psi}_{n_3}(\vec{r}_4)}{E_i+\omega_1-E_{n_3}(1-i0)}
\left(\gamma^{\mu_4} A_{\mu_4}^*(\vec{r}_4)\right)\psi_i(\vec{r}_4).
\end{eqnarray}

Passing to the dipole approximation for transverse photons and using the relations $S_{fi} = -2\pi i U_{fi}\delta(\sum\limits_k E_{fk}-\sum\limits_k E_{ik})$ and $- i m\, \omega_{n k} (\vec{r})_{n k} = (\vec{p})_{n k}$ \cite{LabKlim}, in the nonrelativistic limit the scattering amplitude can be approximately written as
\begin{eqnarray}
\label{S2.6}
U_{fi}^{(4\gamma)} \sim 
\sum\limits_{n_1\, n_2\, n_3}\frac{\langle f| \vec{e}^*_4\vec{r}_4| n_1\rangle\langle n_1| \vec{e}^*_3\vec{r}_3| n_2\rangle\langle n_2| \vec{e}_2\vec{r}_2| n_3\rangle\langle n_3| \vec{e}_1\vec{r}_1| i\rangle}{[\Delta E_{f\, n_1}+\omega_4 + i 0)][\Delta E_{f\, n_2}+\omega_3+\omega_4 + i 0)][\Delta E_{i\, n_3}+\omega_1 + i 0)]},
\end{eqnarray}
where the notation $\Delta E_{c\,d}\equiv E_c-E_d$ is introduced and common factor is omitted for brevity. Resonance occurs when one or all of the energy denominators can turn to zero. By designating the corresponding states as $n_1=r$, $n_2=a$, and $n_3=b$, we can characterize them as absorption $i+\gamma\rightarrow r$, the upper emission link in the cascade $r\rightarrow a+\gamma$, subsequent decay $a\rightarrow b+\gamma$ and the lower cascade link $b\rightarrow f+\gamma$.

The denominators turning to zero should be regularized. This can be done within the framework of the QED theory by the method described in \cite{Low}. This is achieved by inserting an infinite series of consecutive self-energy loops. This procedure should be performed for each electron propagation parts in Fig.~\ref{Fig_S2}. As a result, the energy shift to the state $n_i$ is added in the energy denominators. The self-energy insertions into the outer tails in Fig.~\ref{Fig_S2} can be omitted, see \cite{ANDREEV2008135} for details. As the next step of such an evaluation, one should examine the corresponding energy shift as real and imaginary contributions. The real part is out of interest to us (provided that all necessary energy shifts are included in the energy difference $E_{c\, d}$), and the imaginary part plays a principal role for the study of nonresonant effects due to its relation to the level width. Ordinarily, the regularization procedure is replaced by the phenomenological entry of imaginary additions in the form $-\mathrm{i}\Gamma_{n_i}/2$ into the energy denominator. However, the latter should be applied carefully, since in the case of cascade emission for the lower state the width of the upper level should be taken into account. As a result, in our case the sum of the level widths for the states $a$ and $b$ should arise \cite{ANDREEV2008135,Zalialiutdinov_2014}.

Further, we use the resonance approximation in series: a) according to the third energy denominator in Eq.~(\ref{S2.6}), $E_{n_3}=E_r$ and we denote $E_i+\omega_1-E_r=\omega_{\rm NR}$. Then, the energy conservation law expressed by the $\delta$-function in Eq.~(\ref{S2.5}) yields $E_f+\omega_4+\omega_3+\omega_2-E_r-\omega_{\rm NR}=0$. b) In this relation we substitute $\omega_2=E_r-E_a$ as a free parameter (not explicitly involved in the amplitude Eq.~(\ref{S2.6}), and, therefore, one can find that $E_f+\omega_4+\omega_3-E_a-\omega_{\rm NR}=0$ represents the second energy denominator, i.e. it can be replaced by $\omega_{\rm NR}$ similarly to the third. c) Finally, substituting the resonance value for $\omega_3=E_a-E_b$ into the last energy conservation relation again leads to $E_f+\omega_4-E_b\rightarrow\omega_{\rm NR}$. In the frames of this approximation, we get a one-parameter $\omega_{\rm NR}$ problem instead of a multi-parameter one, in which each frequency is matched with a different $\omega_{\rm NR}(\omega_i)$.

Thus, within the resonance approximation we obtain
\begin{eqnarray}
\label{S2.7}
U_{fi}^{(4\gamma)} \sim \frac{A^{*(1\gamma)}_{f b}A^{*(1\gamma)}_{b a}A^{*(1\gamma)}_{a r}A^{(1\gamma)}_{r i}}{[\omega_{\rm NR}-\frac{\mathrm{i}}{2} (\Gamma_b+\Gamma_a)][\omega_{\rm NR}- \frac{\mathrm{i}}{2}(\Gamma_a+\Gamma_r)][\omega_{\rm NR}- \frac{\mathrm{i}}{2} \Gamma_r]}.
\end{eqnarray}
Here the summation over the projections of all angular momenta is assumed, $A^{(1\gamma)}_{a b}$ denotes the dipole matrix elements on the atomic states $c$ and $d$, and we have discarded all remaining contributions in Eq.~(\ref{S2.6}), which include sums over $n_3\neq r$, $n_2\neq a$ and $n_1\neq b$. In these sums, however, there are terms representing the dominant contribution beyond the resonance approximation. 

For the experiment \cite{Science}, it is necessary to consider two measured frequencies $\nu_{1/2}$ corresponding to $E_{4p_{1/2}^{F=1}}-E_{2s_{1/2 }^{F=0}}$ and $\nu_{3/2}\rightarrow E_{4p_{1/2}^{F=1}}-E_{2s_{1/2}^{F=0} }$. To obtain the effect of quantum interference \cite{Jent-NR,Solovyev2020}, the state $4p_{1/2}^{F=1}$ is combined with $4p_{3/2}^{F=1}$ and vise versa. Thus, using the expression~(\ref{S2.7}), we proceed in the same way, organizing the groups for the lower states $a$ and $b$: $r=\{4p_{1/2}^{F=1},4p_{3/2}^{F=1}\}$, $a=\{3s_{1/2},3d_{3/2},3d_{5/2}\}$, $b=\{2p_{1/2},2p_{3/2}\}$. Thus, the four-photon amplitude in the lowest order beyond the resonant approximation can be written as
\begin{eqnarray}
\label{S2.8}
U_{fi}^{(4\gamma)} \sim  \frac{A^{*(1\gamma)}_{f b}A^{*(1\gamma)}_{b a}A^{*(1\gamma)}_{a r}A^{(1\gamma)}_{r i}}{[\omega_{\rm NR}-\frac{\mathrm{i}}{2} (\Gamma_b+\Gamma_a)][\omega_{\rm NR}- \frac{\mathrm{i}}{2}(\Gamma_a+\Gamma_r)][\omega_{\rm NR}- \frac{\mathrm{i}}{2} \Gamma_r]} + 
\\
\nonumber
\frac{A^{*(1\gamma)}_{f b}A^{*(1\gamma)}_{b a}A^{*(1\gamma)}_{a r'}A^{(1\gamma)}_{r' i}}{[\omega_{\rm NR}-\frac{\mathrm{i}}{2} (\Gamma_b+\Gamma_a)][\omega_{\rm NR}- \frac{\mathrm{i}}{2}(\Gamma_a+\Gamma_{r'})][\omega_{\rm NR} + \Delta E_{r\,r'}- \frac{\mathrm{i}}{2} \Gamma_{r'}]}+
\\
\nonumber
\frac{A^{*(1\gamma)}_{f b}A^{*(1\gamma)}_{b a'}A^{*(1\gamma)}_{a' r}A^{(1\gamma)}_{r i}}{[\omega_{\rm NR}-\frac{\mathrm{i}}{2} (\Gamma_b+\Gamma_{a'})][\omega_{\rm NR}+\Delta E_{a\, a'}- \frac{\mathrm{i}}{2}(\Gamma_{a'}+\Gamma_r)][\omega_{\rm NR}- \frac{\mathrm{i}}{2} \Gamma_r]} + 
\\
\nonumber
\frac{A^{*(1\gamma)}_{f b'}A^{*(1\gamma)}_{b' a}A^{*(1\gamma)}_{a r}A^{(1\gamma)}_{r i}}{[\omega_{\rm NR}+\Delta E_{b\, b'}-\frac{\mathrm{i}}{2} (\Gamma_{b'}+\Gamma_a)][\omega_{\rm NR}- \frac{\mathrm{i}}{2}(\Gamma_a+\Gamma_r)][\omega_{\rm NR}- \frac{\mathrm{i}}{2} \Gamma_r]} + \dots,
\end{eqnarray}
where $\dots$ denote all other terms with different combinations of state groups $r$, $a$ and $b$. Finally, a cross section of the photon scattering process can be obtained by squaring the modulus of amplitude Eq.~(\ref{S2.8}):
\begin{eqnarray}
\label{S2.9}
\sigma_{if} \sim \sum_{M_{F_i}M_{F_f}} \left|\sum_{M}U_{fi}^{(4\gamma)}\right|^2.
\end{eqnarray}
Here we have summed over atomic angular momentum projections in the final state and averaged over the atomic angular momentum projections of initial state, and the sum over $M$ includes all necessary projections.

The transition frequency can be determined from the cross section $\sigma_{if}$ in various ways. One obvious way is to define the transition frequency from the extremum condition as $\omega_{\rm max}$, where $\omega_{\rm max}$ represents the value at which the cross section reaches a maximum. Thus, $\omega_{\rm max}$ is the most probable value. In conjunction with our notation $\omega_{\rm NR} = E_i+\omega_1-E_r\equiv \omega_1-\omega_0$ this condition consists of
\begin{eqnarray}
\label{S2.10}
\frac{d\sigma_{if}}{d\omega_{\rm NR}}=0.
\end{eqnarray}
Employing condition Eq.~(\ref{S2.10}) to the cross section \ref{S2.9} within the resonance approximation, it is enough to consider only the first term in Eq.~(\ref{S2.8}). Then, we immediately arrive at
\begin{eqnarray}
\label{S2.11}
\omega_{\rm max}=\omega_0,
\end{eqnarray}
where, henceforth, $\omega_0$ assumes that all relativistic, QED, etc. corrections are included in the binding energies.

In the lowest order beyond the resonant approximation, this gives
\begin{eqnarray}
\label{S2.12}
\omega_{\rm max}=\omega_0+\delta_{\rm NR}.
\end{eqnarray}
As long as the line profile remains symmetric with respect to $\omega_0$, the definition (\ref{S2.10}) remains the same for any other way of extracting transition frequency from the line profile. For example, for the symmetric line profile the determination of the "line center" \cite{Science} and "line maximum" coincide. The discussion of these two concepts is presented in the main text of the work.

Following equation~(\ref{S2.12}), the nonresonant correction $\delta_{\rm NR}$ can be regarded as a frequency shift. This is valid for cases where the asymmetry of the line profile is small. In mathematical sense, this means that the profile distortion should not exceed its width. A more stringent constraint arises as this correction is calculated, it reads $\Gamma_r/\Delta E_{c\, d}\ll 1$, see \cite{ANDREEV2008135,ZALIALIUTDINOV20181}. After some cumbersome computations, the lowest order correction can be expressed as
\begin{eqnarray}
\label{S2.13}
\delta\omega_i = \frac{f^{(c)}_{\rm nr}}{f^{(c)}_{\rm res}}\frac{\Gamma^2_r}{4\Delta_i}\Upsilon+\dots,
\\
\nonumber
\Upsilon=\frac{(\Gamma_r+\Gamma_a)^2(\Gamma_a+\Gamma_b)^2}{(\Gamma_r+\Gamma_a)^2(\Gamma_a+\Gamma_b)^2+(\Gamma_r+\Gamma_a)^2\Gamma_r^2+(\Gamma_a+\Gamma_b)^2\Gamma_r^2}
\end{eqnarray}
Here $\dots$ denotes corrections of the next orders of magnitude $\Gamma_{r,a,b}$ or coproducts of $\Gamma_r$, $\Gamma_a$ and $\Gamma_b$ divided by $\Delta_i^3$. In contrast to the QIE corrections, the form of Eq.~(\ref{S2.13}) does not permit further simplification (i.e., as a $\Gamma^2_i/\Delta_i$ ratio), since the order of the any widths is almost the same. The most important point, however, is that the ratio of the nonresonant and resonant amplitudes has been distinguished by the multiplier: $f^{(c)}_{\rm nr}/f^{(c)}_{\rm res}$. It is this ratio that determines the angular dependence, and the remaining factor can easily be calculated numerically. 

To obtain an explicit dependence on the angle between the polarization of the incoming photon and the directions of the outgoing photons, it is necessary to calculate the resonant and nonresonant amplitudes included in the expression~(\ref{S2.13}). For this, we turn to the photon scattering process used in \cite{Science} to determine the transition frequency $\nu_{1/2}$ and first determine the associated quantities:
\begin{eqnarray}
\label{S2.14}
f^{(c)}_{\rm res} = \sum\limits_{M_i\,M_f}\left|\sum\limits_{M} A^{*(1\gamma)}_{1s_{1/2} 2p_{1/2}}A^{*(1\gamma)}_{2p_{1/2} 3s_{1/2}}A^{*(1\gamma)}_{3s_{1/2} 4p_{1/2}^{F=1}}A^{(1\gamma)}_{4p_{1/2}^{F=1} 2s_{1/2}^{F=0}} \right|^2,
\end{eqnarray}
\begin{eqnarray}
\label{S2.14a}
f^{(c)}_{{\rm nr}, rr'} = \sum\limits_{M_i\,M_f}\sum\limits_{M} 
A^{*(1\gamma)}_{1s_{1/2} 2p_{1/2}}A^{*(1\gamma)}_{2p_{1/2} 3s_{1/2}}A^{*(1\gamma)}_{3s_{1/2} 4p_{1/2}^{F=1}}A^{(1\gamma)}_{4p_{1/2}^{F=1} 2s_{1/2}^{F=0}}\times
\\
\nonumber
A^{(1\gamma)}_{2s_{1/2}^{F=0} 4p_{3/2}^{F=1}}A^{*(1\gamma)}_{4p_{3/2}^{F=1} 3s_{1/2}}A^{*(1\gamma)}_{3s_{1/2} 2p_{1/2}}A^{*(1\gamma)}_{2p_{1/2} 1s_{1/2}},
\end{eqnarray}
\begin{eqnarray}
\label{S2.14b}
f^{(c)}_{{\rm nr}, aa'} = \sum\limits_{M_i\,M_f}\sum\limits_{M} 
A^{*(1\gamma)}_{1s_{1/2} 2p_{1/2}}A^{*(1\gamma)}_{2p_{1/2} 3s_{1/2}}A^{*(1\gamma)}_{3s_{1/2} 4p_{1/2}^{F=1}}A^{(1\gamma)}_{4p_{1/2}^{F=1} 2s_{1/2}^{F=0}}\times
\\
\nonumber
A^{(1\gamma)}_{2s_{1/2}^{F=0} 4p_{1/2}^{F=1}}A^{*(1\gamma)}_{4p_{1/2}^{F=1} 3d_{3/2}}A^{*(1\gamma)}_{3d_{3/2} 2p_{1/2}}A^{*(1\gamma)}_{2p_{1/2} 1s_{1/2}},
\end{eqnarray}
\begin{eqnarray}
\label{S2.14c}
f^{(c)}_{{\rm nr}, bb'} = \sum\limits_{M_i\,M_f}\sum\limits_{M} 
A^{*(1\gamma)}_{1s_{1/2} 2p_{1/2}}A^{*(1\gamma)}_{2p_{1/2} 3s_{1/2}}A^{*(1\gamma)}_{3s_{1/2} 4p_{1/2}^{F=1}}A^{(1\gamma)}_{4p_{1/2}^{F=1} 2s_{1/2}^{F=0}}\times
\\
\nonumber
A^{(1\gamma)}_{2s_{1/2}^{F=0} 4p_{1/2}^{F=1}}A^{*(1\gamma)}_{4p_{1/2}^{F=1} 3s_{1/2}}A^{*(1\gamma)}_{3s_{1/2} 2p_{3/2}}A^{*(1\gamma)}_{2p_{3/2} 1s_{1/2}}.
\end{eqnarray}
Here it should be noted that decay to the $3d_{5/2}$ state is inherently absent due to the selection rules for the electric dipole transition $4p_{1/2}^{F=1}\rightarrow 3d_{5/2}$.

\begin{figure}[hbtp]
\caption{The QIEc for the transitions $2s^{F=0}_{1/2}\rightarrow 4p^{F=1}_{1/2}(4p^{F=1}_{3/2})\rightarrow 3s_{1/2}\rightarrow 2p_{1/2}(2p_{3/2}),3p_{1/2}(3p_{3/2})\rightarrow 1s_{1/2}$ as the function of the angle between the vectors $\vec{e}_i$, $\vec{\nu}_f$ given by the expressions in Eq.~(\ref{6}). In parentheses the state leading to interference is indicated. The total contribution, $\delta\omega_{\Sigma}$, expressed by the sum of partial channels, is represented by the solid line. The values are plotted in Hz. The effect corresponds to the transition frequency $\nu_{1/2}$ measured in \cite{Science}.}
\centering
\includegraphics[width=0.9\textwidth]{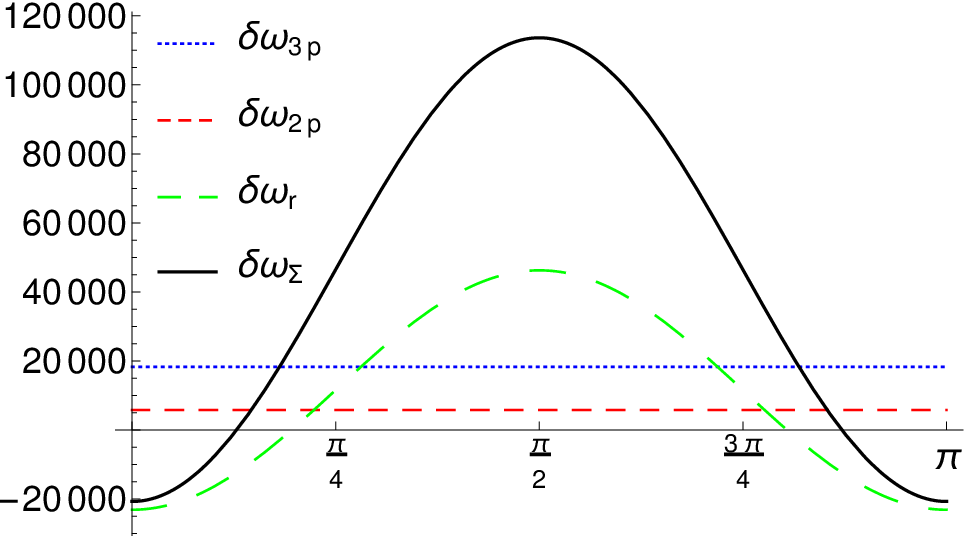}
\label{fig2s}
\end{figure}

\begin{figure}[hbtp]
\caption{The QIEc for the transitions $2s^{F=0}_{1/2}\rightarrow 4p^{F=1}_{1/2}(4p^{F=1}_{3/2})\rightarrow 3d_{3/2}\rightarrow 2p_{1/2}(2p_{3/2}),3p_{1/2}(3p_{3/2})\rightarrow 1s_{1/2}$ as the function of the angle between the vectors $\vec{e}_i$, $\vec{\nu}_f$ given by the expressions in Eq.~(\ref{7}). The other notations are the same as in Fig.~\ref{fig2s}. The values are plotted in Hz.}
\includegraphics[width=0.9\textwidth]{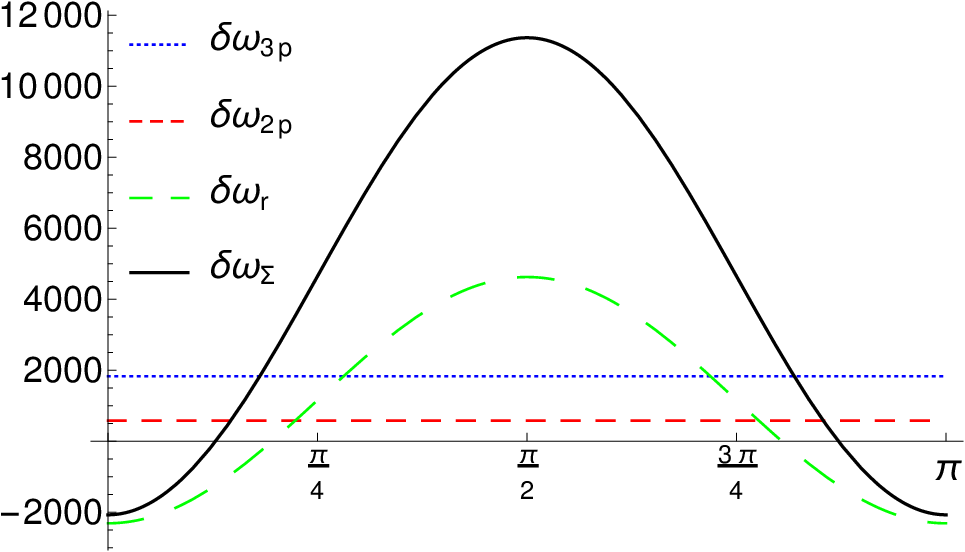}
\label{fig3s}
\end{figure}

\begin{figure}[hbtp]
\caption{The QIEc for the transitions $2s^{F=0}_{1/2}\rightarrow 4p^{F=1}_{1/2}(4p^{F=1}_{3/2})\rightarrow 3s_{1/2}\rightarrow 2p_{3/2}(2p_{1/2}) \rightarrow 1s_{1/2}$ as the function of the angle between the vectors $\vec{e}_i$, $\vec{\nu}_f$. Note that the $3p_{3/2}$ state lies above $3s_{1/2}$, and hence there is no spontaneous decay to this atomic level. The other notations are the same as in Fig.~\ref{fig2s}. The values are plotted in Hz. }
\includegraphics[width=0.9\textwidth]{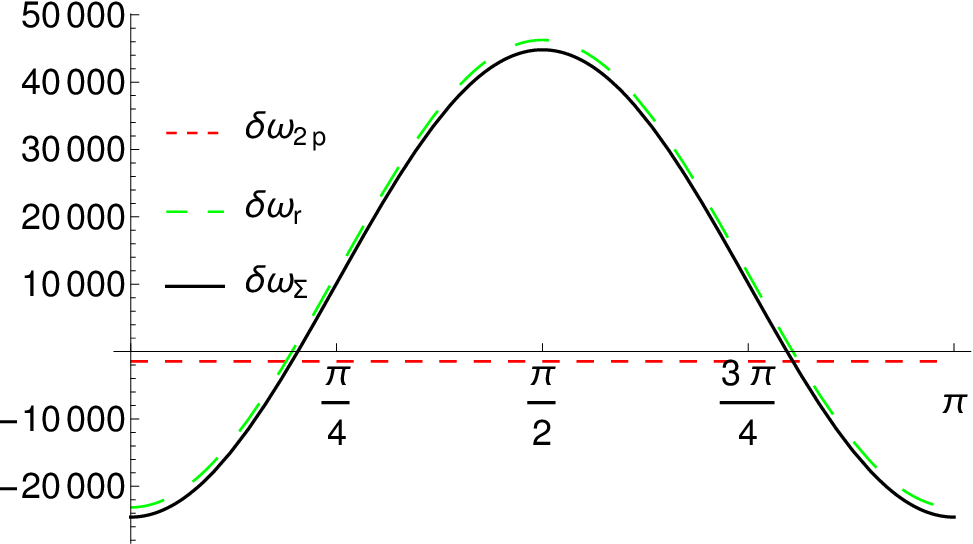}
\label{fig4s}
\end{figure}

\begin{figure}[hbtp]
\caption{The QIEc for the transitions $2s^{F=0}_{1/2}\rightarrow 4p^{F=1}_{1/2}(4p^{F=1}_{3/2})\rightarrow 3d_{3/2}\rightarrow 2p_{3/2}(2p_{1/2}),3p_{3/2}(3p_{1/2})\rightarrow 1s_{1/2}$ as the function of the angle between the vectors $\vec{e}_i$, $\vec{\nu}_f$. The other notations are the same as in Fig.~\ref{fig2s}. The values are plotted in Hz.}
\includegraphics[width=0.9\textwidth]{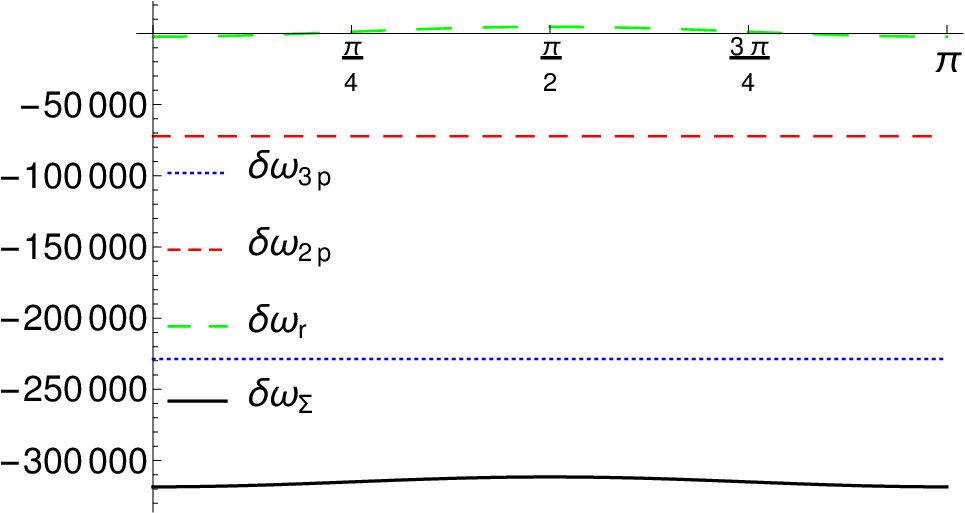}
\label{fig5s}
\end{figure}

The final results for the nonresonant corrections due to QIEc can be demonstrated as graphs in Figs.~\ref{fig2s}-\ref{fig5s}. Numerical results for specific angles are given in Tables~\ref{tab:2}, \ref{tab:3} in the main text and evaluation of matrix elements presented in the next section.

\subsection{Evaluation of matrix elements}
\label{me}
To evaluate matrix elements presented by $A^{(1\gamma)}_{c\, d}$, we use the following relation for scalar product in the cyclic components of two arbitrary vectors, $\vec{w}$ and $\vec{v}$:
\begin{eqnarray}
\label{S2.15}
\vec{w}\vec{v} = \sum\limits_{q=0,\pm 1}(-1)^q w_q v_{-q}.
\end{eqnarray}
The irreducible tensor product of two polarization vectors $\vec{e}_1$ and $\vec{e}_2$ can be expressed as follows:
\begin{eqnarray}
\label{S2.16}
\left\lbrace \vec{e}^{*}_1\otimes\vec{e}_2 \right\rbrace_{x\xi} = (-1)^\xi\Pi_x\sum\limits_{q_1 q_2} 
\begin{pmatrix}
1 & 1 & x\\
q_1 & q_2 & -\xi
\end{pmatrix}
(\vec{e}^{*}_1)_{q1}(\vec{e}_2)_{q_2},
\end{eqnarray}
where $\Pi_{a\,b\,c\dots}=\sqrt{(2a+1)(2b+1)(2c+1)\dots}$, and the standard $3j$-symbol notation is used.

To characterize the atomic state further on, we use a set of quantum numbers $nljFM_F$, where $n$ is the principal quantum number, $l$ is the electron orbital momentum, $j$ is the total angular momentum of the electron, $\vec{j}=\vec{l}+\vec{s}$ ($\vec{s}$ is the electron spin), $F$ is the total atomic momentum, $\vec{F}=\vec{j}+\vec{I}$ ($\vec{I}$ is the nuclear spin) and $M_F$ is the projection of the total atomic momentum. The matrix element of the cyclic component of radius vector is given by \cite{VMK}
\begin{eqnarray}
\label{S2.17}
\langle n'l'j'F'M_{F'}|r_{q}|nljFM_{F}\rangle=(-1)^{F'-M_{F'}}
\begin{pmatrix}
     F' & 1 & F \\
-M_{F'} & q & M_{F}
\end{pmatrix}
\langle n'l'j'F'||r||nljF\rangle,
\end{eqnarray}
where the reduced matrix element is
\begin{eqnarray}
\label{S2.18}
\langle n'l'j'F'||r||nljF\rangle= (-1)^{j'+j+I+l'+1/2+F}
\Pi_{j'jF'F}
\begin{Bmatrix}
j' & F' & I \\
F  & j  & 1
\end{Bmatrix}
\begin{Bmatrix}
l' & j' & 1/2 \\
j  & l  & 1
\end{Bmatrix}
\langle n' l' || r|| nl \rangle\qquad
\end{eqnarray}
(above the ordinary notation for the $6j$-symbol is used), and
\begin{eqnarray}
\label{S2.19}
\langle n' l' || r || nl \rangle = (-1)^{l'}\Pi_{l'l}
\begin{pmatrix}
l & 1 & l'\\
0 & 0 & 0
\end{pmatrix}
\int_{0}^{\infty}r^3 R_{n'l'}R_{nl}dr.
\end{eqnarray}
In Eq.~(\ref{S2.19}) $ R_{nl} $ denotes the radial part of hydrogen wave function. 

Thus, to calculate the squares of the amplitudes in Eqs.~(\ref{S2.14})-(\ref{S2.14c}), we need to sum over the projections of the eight $3j$-symbols according to Eq.~(\ref{S2.17}). The calculations are greatly simplified if we consider that for the initial state $F_i=0$, $M_{F_i}=0$. Then, the relations can be used
\begin{eqnarray}
\label{S2.20}
\begin{pmatrix}
j_1 & j_2 & 0\\
m_1 & m_2 & 0
\end{pmatrix}
 = 
\begin{pmatrix}
0 & j_2 & j_1\\
0 & m_2 & m_1
\end{pmatrix}
=\frac{(-1)^{j_1-m_1}}{\sqrt{2j_1+1}}\delta_{j_1\, j_2}\delta_{m_1\, -m_2}.
\end{eqnarray}

In general, we can arrange a formula convenient for evaluating any amplitudes in Eqs.~(\ref{S2.14})-(\ref{S2.14c}) as follows
\begin{eqnarray}
\label{S2.21}
f^{(c)}_{\rm res/nr} = \sum\limits_{M}(-1)^\phi e_{1_{q_1}}e^*_{2_{q_2}}e^*_{3_{q_3}}e^*_{4_{q_4}} e^*_{1_{p_1}}e_{2_{p_2}}e_{3_{q_3}}e_{4_{q_4}}\times R \times \qquad
\\
\nonumber
\begin{pmatrix}
F_f & 1 & F_b\\
M_{F_f} & -q_4 & M_{F_b}
\end{pmatrix}
\begin{pmatrix}
F_b & 1 & F_a\\
-M_{F_b} & -q_3 & M_{F_a}
\end{pmatrix}
\begin{pmatrix}
F_a & 1 & F_r\\
-M_{F_a} & -q_2 & M_{F_r}
\end{pmatrix}
\begin{pmatrix}
F_r & 1 & F_i\\
-M_{F_r} & -q_1 & M_{F_i}
\end{pmatrix}\times
\\
\nonumber
\begin{pmatrix}
F_i & 1 & F_{r'}\\
M_{F_i} & -p_1 & M_{F_{r'}}
\end{pmatrix}
\begin{pmatrix}
F_{r'} & 1 & F_{a'}\\
-M_{F_{r'}} & -p_2 & M_{F_{a'}}
\end{pmatrix}
\begin{pmatrix}
F_{a'} & 1 & F_{r'}\\
-M_{F_{a'}} & -p_3 & M_{F_{r'}}
\end{pmatrix}
\begin{pmatrix}
F_{r'} & 1 & F_f\\
-M_{F_{r'}} & -p_4 & M_{F_f}
\end{pmatrix},
\end{eqnarray}
where sum over $M$ means the summation over $q_1$, $q_2$, $q_3$, $q_4$, $p_1$, $p_2$, $p_3$, $p_4$, $M_{F_i}$, $F_fM_{F_f}$, $M_{F_r}$, $F_aM_{F_a}$, $F_bM_{F_b}$, $M_{F_{r'}}$, $F_{a'}M_{F_{a'}}$, $F_{b'}M_{F_{b'}}$, the phase corresponds to $\phi = q_1+q_2+q_3+q_4+p_1+p_2+p_3+p_4+F_f-M_{F_f}+F_b-M_{F_b}+F_a-M_{F_a}+F_r-M_{F_r}+F_i-M_{F_i}+F_{r'}-M_{F_{r'}}+F_{a'}-M_{F_{a'}}+F_{b'}-M_{F_{b'}}$. The notation $R$ in Eq.~(\ref{S2.21}) means
\begin{eqnarray}
\label{S2.22}
R = \langle n_fl_f j_f F_f||r||n_b l_b j_b F_b\rangle \langle n_b l_b j_b F_b||r||n_a l_a j_a F_a\rangle \langle n_a l_a j_a F_a||r||n_r l_r j_r F_r\rangle \times
\\
\nonumber
\langle n_r l_r j_r F_r||r||n_i l_i j_i F_i\rangle\langle n_il_i j_i F_i||r||n_{r'} l_{r'} j_{r'} F_{r'}\rangle \langle n_{r'} l_{r'} j_{r'} F_{r'}||r||n_{a'} l_{a'} j_{a'} F_{a'}\rangle \times
\\
\nonumber
\langle n_{a'} l_{a'} j_{a'} F_{a'}||r||n_{b'} l_{b'} j_{b'} F_{b'}\rangle \langle n_{b'} l_{b'} j_{b'} F_{b'}||r||n_f l_f j_f F_f\rangle.
\end{eqnarray}
$R$ is then counted numerically. According to Eq.~(\ref{S2.21}) the resonant or nonresonant squared amplitudes correspond to the choice of the total angular momenta $\{j_r,j_{r'}\}$, $\{j_a,j_{a'}\}$ and $\{j_b,j_{b'}\}$ (equal or not).

Then, applying the expression (\ref{S2.20}) to the $3j$-symbols containing the pair $F_iM_{F_i}$ and the relation (see Eq.~5 in Section 12.1 \cite{VMK})
\begin{eqnarray}
\label{S2.23}
\sum\limits_\kappa(-1)^{q-\kappa}
\begin{pmatrix}
a & b & q\\
\alpha & \beta & -\kappa
\end{pmatrix}
\begin{pmatrix}
q & d & c\\
\kappa & \delta & \gamma
\end{pmatrix}
=
\\
\nonumber
(-1)^{2a}\sum\limits_{x\xi}(-1)^{x-\xi}\Pi^2_x
\begin{pmatrix}
a & c & x\\
\alpha & \gamma & -\xi
\end{pmatrix}
\begin{pmatrix}
x & d & b\\
\xi & \delta & \beta
\end{pmatrix}
\begin{Bmatrix}
b & d & x \\
c  & a  & q
\end{Bmatrix}.
\end{eqnarray}
Thus, we arrive at
\begin{eqnarray}
\label{S2.24}
f^{(c)}_{\rm res/nr} = \sum\limits_{M'}\frac{(-1)^{\phi'}}{3}\Pi^2_{abc}
\begin{pmatrix}
b & 1 & 1\\
\beta & -q_3 & -q_4
\end{pmatrix}
\begin{pmatrix}
a & 1 & 1\\
\alpha & -p_4 & -p_3
\end{pmatrix}
\begin{pmatrix}
F_a & 1 & 1\\
-M_{F_a} & -q_2 & -q_1
\end{pmatrix}
\\
\nonumber
\times\begin{pmatrix}
1 & 1 & F_{a'}\\
-p_1 & -p_2 & M_{F_{a'}}
\end{pmatrix}
\begin{pmatrix}
a & b & c\\
-\alpha & -\beta & -\gamma
\end{pmatrix}
\begin{pmatrix}
c & F_a & F_{a'}\\
\gamma & M_{F_a} & -M_{F_{a'}}
\end{pmatrix}
R\times
\\
\nonumber
\begin{Bmatrix}
1 & 1 & b \\
F_a  & F_f & F_b
\end{Bmatrix}
\begin{Bmatrix}
1 & 1 & a \\
F_f  & F_{a'} & F_{b'}
\end{Bmatrix}
\begin{Bmatrix}
F_{a'} & F_a & c \\
b  & a  & F_f
\end{Bmatrix},
\end{eqnarray}
where summation runs over $q_1$, $q_2$, $q_3$, $q_4$, $p_1$, $p_2$, $p_3$, $p_4$, $F_aM_{F_a}$, $F_b$, $F_{a'}M_{F_{a'}}$, $F_{b'}$, $F_f$, $a\alpha$, $b\beta$ and $c\gamma$. The phase is given by $\phi'=q_1+q_2+q_3+q_4+p_1+p_2+p_3+p_4+F_a-M_{F_a}+F_{a'}-M_{F_{a'}}+a-\alpha+b-\beta+c-\gamma$.

Forming tensor products of polarization vectors as per the expression~(\ref{S2.16}), the final result can be represented in the form:
\begin{eqnarray}
\label{S2.25}
f^{(c)}_{\rm res/nr} = \frac{1}{3}\sum\limits_{F_f F_a F_b}\sum\limits_{F_{a'} F_{b'}}\sum\limits_{abc}\sqrt{\frac{(2a+1)(2b+1)}{(2F_a+1)(2F_{a'}+1)}}(-1)^{b-F_a}R\times
\\
\nonumber
\begin{Bmatrix}
1 & 1 & b \\
F_a  & F_f & F_b
\end{Bmatrix}
\begin{Bmatrix}
1 & 1 & a \\
F_f  & F_{a'} & F_{b'}
\end{Bmatrix}
\begin{Bmatrix}
F_{a'} & F_a & c \\
b  & a  & F_f
\end{Bmatrix}\times
\\
\nonumber
\left(\Big\lbrace\lbrace \vec{e}_3\otimes\vec{e}_4 \rbrace_{b} \otimes \lbrace \vec{e^*}_3\otimes\vec{e^*}_4 \rbrace_{a}\Big\rbrace_c \cdot \Big\lbrace \lbrace \vec{e}_1\otimes\vec{e^*}_2 \rbrace_{F_a} \otimes \lbrace \vec{e^*}_1\otimes\vec{e}_2 \rbrace_{F_{a'}}\Big\rbrace_c\right),
\end{eqnarray}
where $\cdot$ means the scalar product.

Further evaluation of Eq.~(\ref{S2.25}) is performed numerically. In such calculations, for each squared amplitude we sum over the total atomic momenta $F_n$ for the final and intermediate states with the specified total angular momenta. Then, the nonezero tensor components summed over the polarizations $\vec{e}_2$, $\vec{e}_3$, $\vec{e}_4$ and in the assumption that the emitted photons are detected in one direction (i.e., $\vec{\nu}_2\parallel\vec{\nu}_3\parallel \vec{\nu}_4$) can be found as
\begin{eqnarray}
\label{S2.26}
\sum\limits_{\vec{e}_3\vec{e}_4}\Big\lbrace\lbrace \vec{e}_3\otimes\vec{e}_4 \rbrace_0 \otimes \lbrace \vec{e^*}_3\otimes\vec{e^*}_4 \rbrace_0\Big\rbrace_0 = \frac{1}{3},
\\
\nonumber
\sum\limits_{\vec{e}_3\vec{e}_4}\Big\lbrace\lbrace \vec{e}_3\otimes\vec{e}_4 \rbrace_2 \otimes \lbrace \vec{e^*}_3\otimes\vec{e^*}_4 \rbrace_2\Big\rbrace_0 = \frac{2}{3\sqrt{5}},
\\
\nonumber
\end{eqnarray}
and
\begin{eqnarray}
\label{S2.27}
\sum\limits_{\vec{e}_2}\Big\lbrace\lbrace \vec{e}_1\otimes\vec{e^*}_2 \rbrace_0 \otimes \lbrace \vec{e^*}_1\otimes\vec{e}_2 \rbrace_0\Big\rbrace_0 = \frac{1}{3}\sin^2\theta,
\\
\nonumber
\sum\limits_{\vec{e}_2}\Big\lbrace\lbrace \vec{e}_1\otimes\vec{e^*}_2 \rbrace_1 \otimes \lbrace \vec{e^*}_1\otimes\vec{e}_2 \rbrace_1\Big\rbrace_0 = \frac{\cos^2\theta}{2\sqrt{3}},
\\
\nonumber
\sum\limits_{\vec{e}_2}\Big\lbrace\lbrace \vec{e}_1\otimes\vec{e^*}_2 \rbrace_2 \otimes \lbrace \vec{e^*}_1\otimes\vec{e}_2 \rbrace_2\Big\rbrace_0 = \frac{1}{6\sqrt{5}}(3+\sin^2\theta).
\end{eqnarray}

\end{document}